\documentclass[12pt,a4paper,sort&compress]{article}
\usepackage{amsmath, amssymb, amscd, amsthm, amsfonts}
\usepackage{graphicx}
\usepackage{hyperref}
\usepackage{algorithm}
\usepackage{algpseudocode}
\usepackage{comment}
\usepackage{pdfpages}

\usepackage{todonotes}
\usepackage{mathrsfs}

\usepackage{xcolor}

\usepackage[numbers,sort&compress]{natbib}
\usepackage{array}
\usepackage{authblk}
\usepackage{mathtools}
\usepackage{soul}
\usepackage[symbol]{footmisc}
\usepackage[resetlabels]{multibib}

\usepackage{lineno}

\newcites{App}{Appendix References}

\usepackage{booktabs}
\title{\vspace{-4cm}On the Salient Limitations of the Methods of Assembly Theory and their Classification of Molecular Biosignatures}
\author[1,2]{Abicumaran Uthamacumaran}
\author[3,4]{Felipe S. Abrah\~{a}o}
\author[5,6]{Narsis A. Kiani}
\author[ 6,7]{Hector Zenil\thanks{ Corresponding author. Email: hector.zenil@cs.ox.ac.uk, hector.zenil@kcl.ac.uk}}

\affil[1]{\normalsize\text{ } Department of Physics and Psychology (Alumni), Concordia University, Canada. \normalsize}

\affil[2]{ McGill University, McGill Genome Center, Majewski Lab, Canada.}

\affil[3]{ Centre for Logic, Epistemology and the History of Science, University of Campinas (UNICAMP), Brazil. \normalsize}

\affil[4]{ DEXL, National Laboratory for Scientific Computing, Brazil. \normalsize}

\affil[5]{ Department of Oncology-Pathology, Center for Molecular Medicine, Karolinska Institutet, Sweden. \normalsize}

\affil[6]{ Algorithmic Dynamics Lab, Karolinska Institutet, Sweden. \normalsize}

\affil[7]{ School of Biomedical Engineering and Imaging Sciences, King's College London, U.K.. \normalsize}

\date{}

\begin{document}

%% Mathematical environment added by Felipe:

\theoremstyle{definition}
\newtheorem{definition}{Definition}[section]

\newtheorem{lemma}{Lemma}[section]
\newtheorem{theorem}{Theorem}[lemma]
\newtheorem{corollary}{Corollary}[theorem]

        \maketitle
\vspace{-1.3cm}
%\newpage 
\begin{abstract}
	
	%	In a recent paper, a method and measure have been proposed claiming to be capable of identifying molecules as biosignatures with what they call `Assembly Theory' on data from mass spectrometry, and in this manner tell apart living from non-living.  
	%	In this article, we show that the assembly pathway method is a suboptimal restricted version of Huffman's encoding so that it falls into the category of a purely weak entropic measure. 
	%	Having identified a lack of control experiments and a limited analysis offered, we compare other measures of statistical and algorithmic nature that perform similarly, if not better, than the original proposed at identifying such molecular signatures without making recourse to a new theory. 
%	\begin{linenumbers}

We demonstrate that the assembly pathway method underlying assembly theory (AT) is an encoding scheme widely used by popular statistical compression algorithms. We show that in all cases (synthetic or natural) AT performs similarly to other simple coding schemes and underperforms compared to system-related indexes based upon algorithmic probability that take into account statistical repetitions but also the likelihood of other computable patterns. Our results imply that the assembly index does not offer substantial improvements over existing methods, including traditional statistical ones, and imply that the separation between living and non-living compounds following these methods has been reported before.\\

\noindent \textbf{Keywords:} assembly theory, assembly index, complexity, biosignatures, statistical coding, algorithmic information, LZ compression
		
%	\end{linenumbers}
\end{abstract}

%\newpage

%\linenumbers

\section{Introduction}\label{sectionIntro}

The distinction between living and nonliving systems has long fascinated both scientists and philosophers. 
The question has been at the core of the areas of systems biology and complexity science since their inception, while the seminal concept of complexity---an irreducible emergent property among simpler components in a system---has long been believed to be central to the distinction between living systems and inanimate matter ~\cite{Mitchell2009,Walker2012nat,Adams2017,Prokopenko2009,Abrahao2021bpublished}.

The first to discuss this nexus of issues was Erwin Schrödinger, in his book ``What is Life?'', exploring the physical aspect of life and cells, followed by Claude Shannon, whose concept of entropy, significantly shaped not only by communication theory but by his characterisation of life and intelligence,
placed the concept of information at the core of the question about life. Shannon proposed that his digital theory of communication and information be applied to understanding information processing in biological systems~\cite{shannon}. 
%responding to the pressing challenge of identifying the distinctiveness of certain configurations of atoms or molecules assembled non-randomly, and quantifying the many ways in which swapping these molecules could explain a whole system. 

By solving not only the problem of a mathematical definition for randomness but also the apparent bias toward simplicity underlying formal theories, the concepts of algorithmic information, algorithmic randomness, and algorithmic probability from Algorithmic Information Theory (AIT) abstract the issue away from statistics and human personal biases and choices to recast it in terms of fundamental mathematical first principles. 
These foundations are the underpinnings of coding methods, and they are ultimately what explain and justify their application as a generalisation of Shannon's information theory. %which is subsumed into AIT. 
AIT has also been motivated by questions about randomness, complexity, and structure in the real world, formulating concepts ranging from algorithmic probability~\cite{solo}, that formalises the discussion related to how likely a computable process or object is to be produced by chance under information constraints, to the concept of logical depth~\cite{bennett}, that frames the discussion related to process memory, causal structure and how life can be characterised otherwise that in terms of randomness and simplicity.

A recently introduced approach termed ``Assembly Theory" (AT), featuring a computable index, has been claimed to be a novel and superior approach to distinguishing living from non-living systems and gauging the complexity of molecular biosignatures with an assembly index or molecular assembly index (MA).  
%A major problem in science is that of reproducibility and lack of proper control experiments. 
In proposing MA as a new complexity measure that quantifies the minimal number of bond-forming steps needed to construct a molecule, the central claim advanced in~\cite{cronin} is that molecules with high molecular assembly index (MA) values ``are very unlikely to form abiotically, and the probability of abiotic formation goes down as MA increases''. 
In other words, according to the authors, ``high MA molecules cannot form in detectable abundance through random and unconstrained processes, implying that the existence of high MA molecules depends on additional constraints imposed on the process''~\cite{cronin}.
We will use the notation `AT', `assembly index', or `MA' to refer to the aforementioned theory and the index derived therefrom.

%Note that, depending on variations in application and context, the index measure featured in
%Assembly Theory has had several names, or can be referred to in different ways: pathway assembly (PA), object assembly (OA), or molecular assembly index (MA) as in~\cite{Marshall2019main,marshall_murray_cronin_2017main}. 
%We choose to employ the last nomenclature in the present article.
%%, this paper's results and conclusion hold across all their methods and measures based on \cite{Marshall2019main,marshall_murray_cronin_2017main,cronin}.

The underlying intuition is that such an assembly index (by virtue of minimising the length of the path necessary for an extrinsic agent to assemble the object) would afford ``a way
to rank the relative complexity of objects made up of the same building units on the basis of the pathway, exploiting the combinatorial nature of these combinations''~\cite{marshall_murray_cronin_2017}.

In order to support their central claim, the authors of Assembly Theory state that ``MA tracks the specificity of a path through the combinatorially vast chemical space''~\cite{marshall_murray_cronin_2017} and that, as presented in~\citet{Marshall2019}, it ``leads to a measure of structural complexity that accounts for the structure of the object and how it could have been constructed, which is, in all cases, computable and unambiguous''.
%In particular, \citet{Marshall2019} sets the formalism on which \cite{cronin} is based.
%and for this reason we also base our notation and definitions in the following theorems on \cite{Marshall2019}.

\subsection{What a ZIP file can tell about life}

The authors propose that molecules with high MA detected in contexts or samples generated by random processes, in which there are minimal (or no) biases in the formation of the objects, display a smaller frequency of occurrence in comparison to the frequency of occurrence of molecules in alternative configurations, where extrinsic agents or a set of biases (such as those brought into play by evolutionary processes) play a significant role.

However, we found that what the authors have called AT~\cite{cronin} is a formulation that mirrors the working of previous coding algorithms---though no proper references or attributions are offered---in particular, statistical lossless compression algorithms, whose purpose is to find redundancies~\cite{ATpaper2}. These algorithms were dictionary-based, like run-length encoding (RLE), Huffman~\cite{huffman_1952}, and Lempev-Ziv (LZ)-based~\cite{lz}. They were all launched early in the development of the field of compression for the purpose of detecting identical copies that could be reused. 

%The algorithms in the LZ family of compression schemes, in particular, look for the longest substring matches. 
Lossless compression, incorporating the basic ideas of LZ compression, has been widely applied in the context of living systems, including in a landmark paper published in 2005, where it was shown that it was not only capable of characterising DNA as a biosignature, but also of reconstructing the main branches of an evolutionary phylogenetic tree from the compressibility ratio of mammalian mtDNA sequences~\cite{licompression}. The same LZ algorithms have been used for plagiarism detection, as measures of language distance, and for clustering and classification~\cite{licompression}.
In genetics, it is widely known that similar species have similar nucleotide GC content, and that therefore a simple Shannon Entropy approach on a uniform distribution of G and C nucleotides---effectively simply counting the exact repetitions of polymers~\cite{zenilgc}---can yield a phylogenetic tree. LZ compression has been used in this same context~\cite{review}, and is central to complexity applications to living organisms, which are based upon exactly the same grounds and on the idea of repetitive modules. 

LZ77/LZ78 is at the core of AT, but its assembly index method is weaker than resource-bounded measures introduced before~\cite{ctm,soler,bdm}. 
LZ-based schemes have been used in compression since 1977, and they are behind algorithms like zip, gzip, giff, and others, exploited for the purposes of compression and as approximations to \emph{algorithmic} (Solomonoff-Kolmogorov-Chaitin) \emph{complexity}, which is one of the indexes from AIT.
This is because compressibility is sufficient proof of non-randomness.
Being one of the LZ compression schemes~\cite{ATpaper2}, the assembly index calculation method looks for the largest substring matches, counting them only once as they can be reused to reproduce the original object. But it is weaker than other approximating measures because by definition it only takes into consideration identical copies rather than the full spectrum of causal operations to which an object may be subject (beyond simple identical copies).

Our results demonstrate that the claim that AT may help not only to distinguish life from non-life but also to identify non-terrestrial life, explain evolution and natural selection, and unify physics and biology is a major overstatement. 
(See also the Appendix for a detailed presentation of the results).
What AT amounts to is a re-purposing of some elementary algorithms in computer science in a sub-optimal application to life detection that has been suggested and undertaken before~\cite{bennett,zenilld}, even generating the same results when applied to separating organic from non-organic chemical compounds~\cite{zenilchem}.
%While the calculation of MA may be prone to false negatives---due to partial fragmentation in energy collision analysis and the restriction to counting only valence rules in molecule synthesis (ignoring other chemical conditions)---this does not pose a challenge to the central claim made in~\cite{cronin}. 
%Instead, MA aims at avoiding underestimation of the amount of molecules that result from random or abiotic processes. Thus in the present article, instead of studying both positives and false negatives, we only focus on investigating the existence of false positives, which directly tackles the central claim. 
%This is consonant with understanding MA as a pathway (or object) assembly complexity measure, as presented in \cite{Marshall2019,marshall_murray_cronin_2017}, and 
%The limitations and drawbacks identified here extend to all applications of these methods developed in~\cite{cronin,Marshall2019,marshall_murray_cronin_2017,croninnature} and are based on their comparison to other weak statistical measures. 
By empirically demonstrating the higher predictive performance of AIT-based complexity measures, such as approximations to algorithmic complexity, to experimental applications in molecular classification, we extend the results reported before in~\cite{zenilchem} that had already---years before the introduction of Assembly Theory---demonstrated the capabilities of these measures as regards separating chemical compounds by their particular properties, including organic from inorganic compounds.
Further research based on the same underlying ideas of perturbation/mutation analysis together with algorithmic information theory has also been recently used to detect and decode bio- and technosignatures~\cite{et}.

\section{MA and compression algorithms}

By employing different types of data (on the same subset of molecules~\cite{marshall_murray_cronin_2017,cronin}), as shown in Figures~\ref{SupInffigureCompressioncorrelationplot} and~\ref{SupInffigureCompressioncorrelationplot2}, we demonstrate that other measures applied to other (chemical and molecular) data reproduce what AT's authors claimed was unique, though in fact it was not. 
We have shown that the same indexes used and shown in these figures, and reported to separate organic from non-organic compounds before in~\cite{zenilchem}, also separate what the authors thought was a unique type of spectral data. 
Using exactly the same data input utilised by the authors of AT in their original paper~\cite{cronin}, we have shown that their MA index, also known as the assembly index, displays exactly the same behaviour as other complexity indexes. 
These results show that the assembly index calculation method not only is a compression scheme (as proven in~\cite{ATpaper2}), but also performs like one for all intents and purposes, and does not seem to afford any classificatory advantage either by virtue of its method or in combination with any property of the input data (e.g. mass spectra).

%% Figure 4

\begin{figure}[ht!]
	\centerline{\includegraphics[scale=0.11]{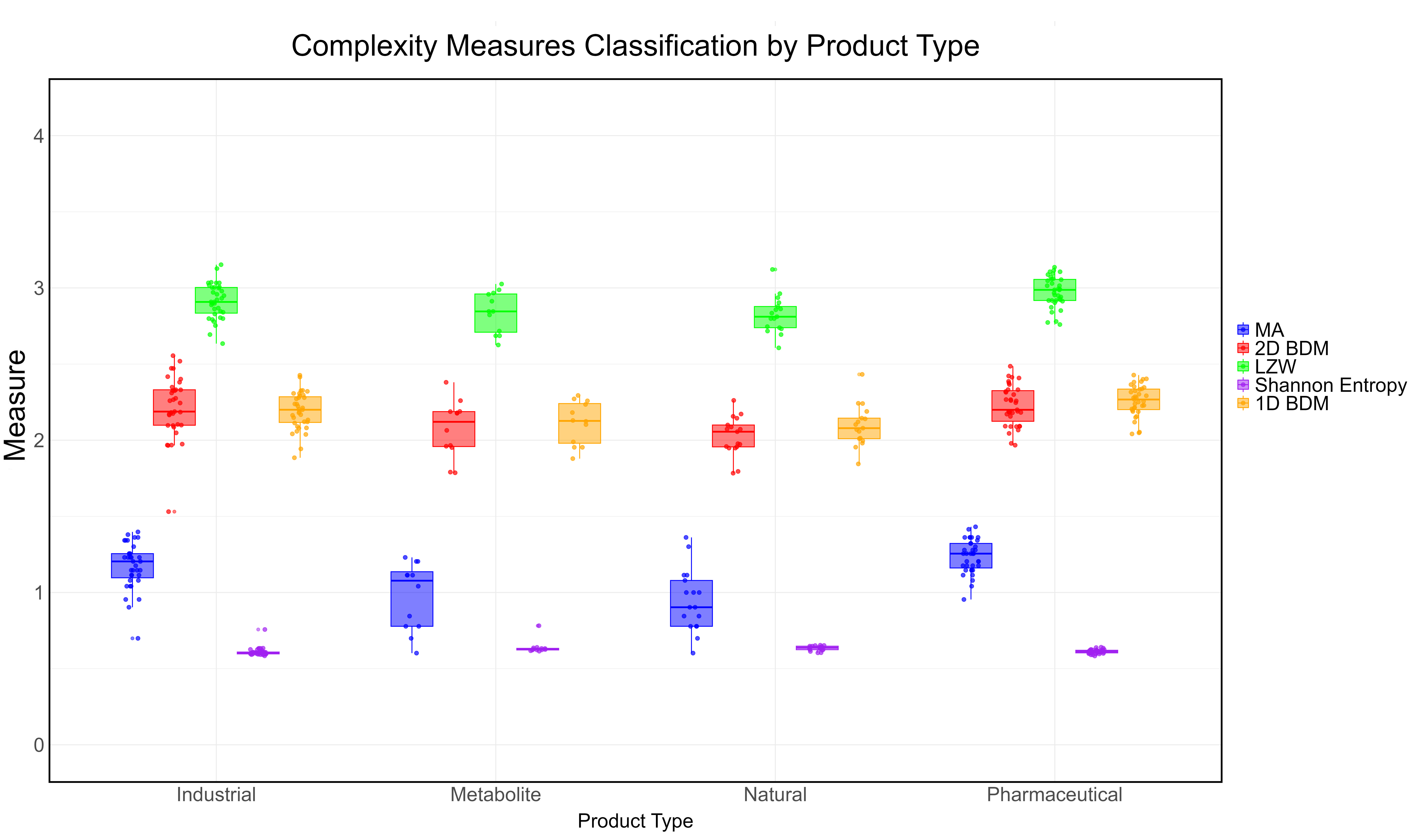}}
	\caption{\label{figureLivingversusNonliving}Classification of molecular complexity by multiple complexity indexes originally used to create the chemical space for the mass spectroscopy (MS) profiles (log-scale). 
	A strong Pearson correlation with an R-value of 0.8823 was observed between 1D-BDM and MA for the 99 molecules available in the MS data set.
	LZW compression shared a close Pearson's correlation score of 0.8738 with MA. All correlation measures obtained a statistically significant one-tailed p-value ($ P < 0.0001 $).
	All measures other than MA applied to bond molecular distance matrices, some of which outperform MA and mass spectra at distinguishing organic from non-organic molecules found in the MS dataset of the MA paper~\cite{cronin}, as demonstrated by greater separation and smaller variance results across the different complexity measures among the molecular subgroups. MA does not display any particular advantage when compared against proper control experiments, and performs similarly to the simplest of the statistical algorithms applied to all the tested data representations, including molecular distance matrices (as shown here for all measures but MA) or the mass spectral data provided by the authors of Assembly Theory (shown on the plot from the authors' results that could not be fully reproduced due to lack of data made available in~\cite{cronin} but which we took at face value) for comparison purposes.}
\end{figure}

\begin{figure}[ht!]
	\centerline{\includegraphics[scale=0.115]{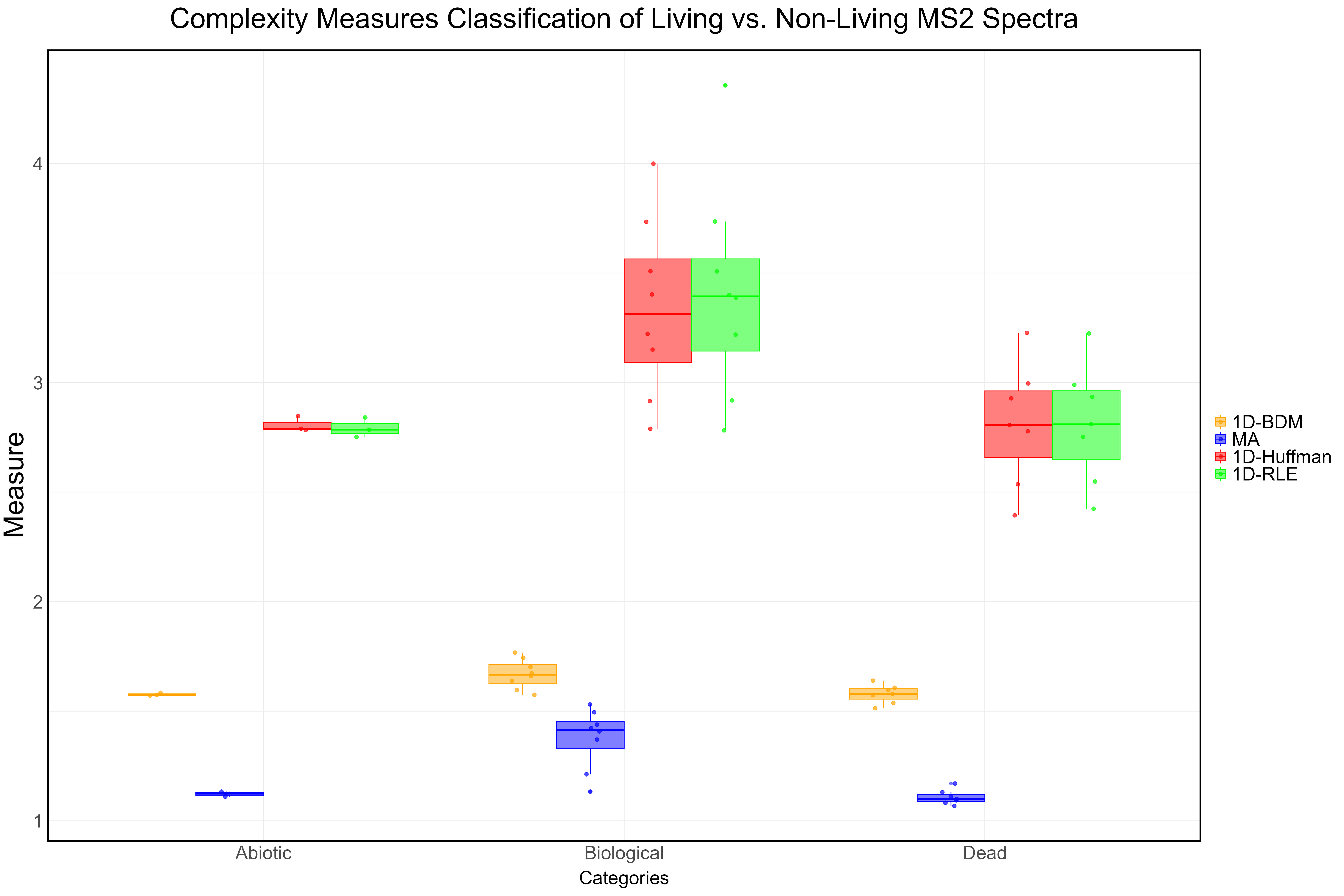}}
	\caption{\label{figureBiologicalComplexityplot}Analysis of organic versus non-organic molecules from mass spectral data by multiple complexity indexes: The strongest positive correlation was identified between MA and 1D-RLE coding (R= 0.9), which is one of the most basic coding schemes and among the most similar to the intended definition of MA, as being capable of `counting copies' in 18 extracts for which the mass spectra was available. Other coding algorithms, including LZ and Huffman coding (R = 0.896), also show a strong positive correlation with MA. As seen, the compression values of both 1D-RLE and 1D-Huffman coding show overlapping and nearly identical medians (horizontal line at centre) and ranges on the whisker plot. The analysis further confirms our previous findings, with the similarity in performance 
	in classifying living \textit{vs}$.$ non-living between MA and popular statistical compression measures (whose purpose is also to count identical statistical copies) leading us to make the case that MA is one 
	(and the same as compression).}
\end{figure}

Assembly Theory claims that MA can predict living \textit{vs}$.$ nonliving molecules, testing it against a small cherry-picked subset of biological extracts, between abiotic factors and inorganic (dead) matter. 
We repeated the experiment using the binarised MS2 spectra peaks matrices provided in the source data in~\cite{cronin}. 
Our reproduced findings are shown in Figures~\ref{figureLivingversusNonliving} and~\ref{figureBiologicalComplexityplot}. 
(See also the Appendix~\ref{sectionSupInf-Results} for more detailed information).
%By including the 114 molecules from Figure 3 with the 18 molecules of their Figure~\ref{figureBiologicalComplexityplot}, we performed correlation analysis on all 132 signatures, with 5 categories: small molecules, peptides, abiotic, dead (inorganic, such as coal and quartz) and biological extracts (which includes yeast, E.coli, etc.). 
%The Pearson correlation was strongest between 1D-BDM and the category (R= 0.951), followed by 1D-RLE and 1D-Huffman having a near-identical Pearson correlation of R= 0.843 and R= 0.842, respectively. MA has the poorest correlation with the categories, with a correlation of R= 0.448. All Pearson scores were statistically significant (P$<$0.0001). 
%The results for all 132 molecules are shown in Table 3. 

%%%

%This conforms with the expectation of the index proven to be, mathematically speaking, a weak approximation to algorithmic complexity indistinguishable from Shannon Entropy and from statistical compression such as LZ77/LZ78~\cite{ATpaper2}. This is because 

Thus, the coding indexes systematically outperform the MA index as a discriminant of living \textit{vs}$.$ non-living systems.
MA works on the basis upon which all popular statistical lossless compression algorithms operate, the principle of `counting exact repetitions' in data, which AT fully relies upon. These are basic coding schemes introduced at the inception of information theory and computer science that do not incorporate the many advances made in recent decades in the area of coding, compression, and resource-bounded algorithmic complexity theory~\cite{aidbook} and cannot explain selection and evolution or unify physics and biology~\cite{croninnature} beyond the connections already made~\cite{santiago}. 

As demonstrated here, the characterisation of molecules using mass spectrometry signatures is not a challenge for other equally computable and statistically-driven indexes.
Other indexes are equally capable of discriminating biosignature categories, by InChI, by bond distance matrices or by mass spectra (MS2 peak matrices), thus disproving the claim that MA is the only experimentally valid measure of molecular complexity.

\section{Limitations of MA as a complexity measure}

We have also shown that as soon as the MA index is confronted with more complicated cases of non-linear modularity, it underperforms or misses obvious regularities. 
As shown in this article and more detailed in the Appendix, our results show that MA, and its generalisation in the hypothesis called AT, is prone to false positives and fails both in theory and in practice to capture the notion of high-level causality beyond non-trivial statistical repetitions---that Shannon Entropy could not have already captured in the first place---which is necessary for distinguishing a serendipitous extrinsic agent (e.g. a chemical reaction resulting from biological processes) that constructs or generates the molecule of interest from a simple or randomly generated configuration (e.g. a chemical reaction resulting from environmental catalytic processes) or crystal-like minerals, as corroborated in~\cite{royalsocietyminerals}.

The statistically significant separation of organic from non-organic compounds using molecular data and approximations to algorithmic complexity via compression, including using structural distance matrices empirically not very different from the mass spectral data used by AT, was first reported in~\cite{zenilchem}. 
In another paper, we also made connections to selection and evolution, predating by several years~\cite{santiago} a recent paper based on the same principles by the same group~\cite{croninnature}, but unlike this paper ours included tests on actual biological data, including but going beyond simple statistical repetitions (exact copies) using a Block Decomposition Method~\cite{bdm}.
Very similar arguments and measures to those set forth by the authors of AT formulated  in~\cite{santiago} show, with actual generative examples, how modularity may emerge from simple mechanistic processes that follow algorithmic probability, explaining what AT meant to explain regarding how evolution may shortcut random processes towards building functional modular and hierarchical systems, and how evolution, drawing from a simplicity-biased distribution imposed by physical and chemical laws would, in a very fundamental fashion, lead to known evolutionary phenomena. 

The present article shows that the authors have failed to cite essential prior literature, mostly rehashing concepts and measures introduced before.
The claims regarding the capabilities of AT---to characterise life, redefine time, find extraterrestrial life, explain selection and evolution, and unify biology and physics~\cite{croninnature}---are shown to be unfounded or exaggerated, and if true, the same would be true of most of these other indexes. 

In summary, while in~\cite{ATpaper2} it is shown that AT is formally equivalent to a compression method (so that the assembly index calculation method is demonstrated to belong to the LZ family of compression schemes), here we have empirically shown that the best performance of molecular assembly does not outdistance other measures of a statistical nature (e.g., those based on Shannon Entropy) in any input data tested.
Therefore, it conforms with the theoretical expectation, and highlights a well-known mathematical property in data compression and complexity science:
specifically, that different parsings (of an object) can perform equally in terms of compression rate. 
This directly reveals that the illustrative examples presented in later work~\cite{Kempes2024ATreply2} fail to address our results; and further attempts in~\cite{Foote2022ATreply3,Walker2024ATreply1} seem to overlook the intrinsic deficiencies (both theoretical and empirical) in AT demonstrated in the present article.

Thus, we do not find AT to make deep or meaningful contributions to advancing the field, or to introducing new concepts, methods, novel applications or results that had not already been introduced or reported, especially in light of the hyperbolic claims associated with Assembly Theory, and its multiple failures to cite the relevant literature.
The limitations and drawbacks identified here extend to all applications of these methods developed in~\cite{cronin,Marshall2019,marshall_murray_cronin_2017,croninnature} and are based on their comparison to other weak statistical measures.

\section{Discussion: emergence and intrinsic complexity measures}

Living systems are complex systems consisting of multiscale, multi-nested processes that are unlikely to be reducible to simplistic and intrinsic statistical properties such as those suggested by AT. 
Pure stochasticity is too strong an assumption and does not realistically represent the generative processes of molecules.
Especially in the context of complex systems like living organisms, organic molecules may be the byproduct of intricate combinations of deterministic/computable and stochastic processes that govern the behaviour of the entire organism~\cite{Walker2012nat,bdm,zenilthermo} and its relationship with the way such agents exploit and interact with the information in their environments. In attempting to determine the living nature of an agent, any complexity measure that only looks at the agent's internal structure, not taking into account an agent's relationship with environmental state variables, is destined to fail. 

We have shown that, lacking the capability of detecting essential features of complex structure formation that go beyond a linear and combinatorial sequence space optimised for statistically identical repetitions, AT and its mathematical and computational methods based on decades-old coding schemes may return misleading values that would classify a low-complexity molecule as being extrinsically constructed by a much more complex agent, thus failing to appropriately characterise extraterrestrial life, contra the authors' claim~\cite{cronin,croninnature}. 
This extrinsic agent may be of a much simpler nature (e.g. a naturally occurring phenomenon).
That is, in case a sufficiently complex environmental catalytic condition plays the role of this extrinsic factor (which increases the bias toward the construction of a more complex molecule), such a level of complexity would be completely missed by the capabilities of a simplistic measure such as MA, thereby rendering it prone to false positives.
%While it may be that the discussed complexity measures and algorithms we presented are a powerful set of tools (framework) to help identify causal patterns, to classify biosignatures from living systems from those of non-living factors, it is one of many paradigms to characterise biological dynamics. 
The presence of emergent properties that characterises the complexity of living systems, cannot be reduced to a single paradigm or dimensionality, further confirming irreducibility as a hallmark of complex systems \cite{zenilthermo,Abrahao2021bpublished}. 

\section*{Data availability}\label{sectionDataavailability}

All the results, data and code are provided in our project GitHub repository:
\url{https://github.com/Abicumaran/MSComplexity/}
Mass spectrometry data is available in the Supplementary Information of~\cite{cronin}.

The Online Algorithmic Complexity Calculator (OACC) to reproduce the values of complexity indexes is available at: \url{http://www.complexity-calculator.com/}.
Text to binary conversion is available at: \url{https://www.rapidtables.com/convert/number/ascii-to-binary.html}.
The results of compression algorithms can be reproduced using:
\url{https://planetcalc.com/9069/}   for the Lempel-Ziv-Welch (LZW);
\url{https://www.dcode.fr/rle-compression} for the run-length encoding (RLE);
\url{https://www.dcode.fr/huffman-tree-compression} for the Huffman Coding.

\subsection*{Code availability}

Statistical correlation analysis was performed using GraphPad Prism v. 8.4.3., available at \url{https://www.graphpad.com/scientific-software/prism/}.
Further computational tools to reproduce our results are described in the section `Data availability'.

\section*{Acknowledgements}

Felipe S. Abrah\~{a}o acknowledges support from the São Paulo Research Foundation (FAPESP), grants $2021$/$14501$-$8$ and $2023$/$05593$-$1$.

%\section*{Author contributions}
%
%Writing: A.U., H.Z. and F.S.A.; experimental design, methods and analysis: A.U. and H.Z.; conceptualisation: H.Z., A.U., and F.S.A.; 
%formal analysis: A.U., F.S.A., and H.Z.; revision and reorganisation: N.A.K;
%supervision: H.Z. All authors have read and agreed to the published version of the manuscript.
%
%\section*{Competing interests}
%
%The Authors declare the absence of any Competing Financial or Non-Financial Interests.

%% Felipe's comment: I just changed the bibliography style in order to use \citet from the natbib package. 
%\bibliographystyle{alpha}
\bibliographystyle{plainnat}
\bibliography{references} % see references.bib for bibliography management

\appendix

\section{Mischaracterisations}
\label{sectionSupInfComputationalaspects}

To understand the mathematical limitations underpinning AT, first note that the pathway assemblages are characterised by functions of the form
%$f_n@(x_1, x_i, x_n)$
\begin{equation*}
	\begin{aligned}
		%$ 
		g_k \colon 
		\setlength\arraycolsep{0pt} 
		\begin{array}[t]{c >{{}}c<{{}} c}
			V\left( \Gamma \right) \times  V\left( \Gamma \right) & \to & V \left( \Gamma \right) \\ 
			\left( z , x  \right) = \left( z , \left(  w_1 , \dots , w_k , \dots \right) \right) & \mapsto & g_k\left( z , x  \right) = \left( w_1 , \dots , f\left( z , w_k \right) , \dots \right)
		\end{array} 
		\text{ ,}
		%$
	\end{aligned}
\end{equation*}
where $ \left(  w_1 , \dots , w_k , \dots \right) $ denotes the object $ x $ in the assembly space $ \left( \Gamma , \phi \right) $ that results from the combination of other objects $ w_1 $, $ w_2 $, $ \dots$, $ w_k $, etc and
function $ \, f \colon V\left( \Gamma \right) \times  V\left( \Gamma \right) \to V\left( \Gamma \right) \, $ gives the result of combining object $ z $ with $ w_k $.
Being limited to joining operations---and this limitation becomes even more dramatic in the generative processes that we will discuss below---AT cannot deal with any variation of $ x $ or $ f $ beyond successive simple constructions.
In the general case, most computable objects would be missed by statistical methods (like entropy and cognates such as AT).
Since probability distribution uniformity does not guarantee randomness~\cite{zenil_kiani_2017,Becher2002,Calude2002main}, most objects, both in theory and practice, cannot be recognised or characterised by weak computable measures, especially by those that are largely based on entropy measures such as statistical compression algorithms or AT.
%This occurs because not every block of data is equally represented so those measures will mischaracterise objects that are highly algorithmic modular in its composing structure~\cite{zenil_kiani_2017}.

Such a mischaracterisation has its roots in the reason any particular statistical test may fail to capture a mathematical formalisation of randomness, an inadequacy which prompted the positing of algorithmic randomness~\cite{Calude2002main,Downey2019}.
For every computable statistical test (e.g., obeying the law of large numbers or displaying Borel normality) for which there is a computably enumerable number of sequences that satisfy it, there are arbitrarily large initial segments of sequences that can be computed by a program, although these initial segments would be deemed random by statistical tests.

On the contrary, algorithmic randomness requires the sequence to be incompressible (and, as a consequence, uncomputable) across the board, or to pass \emph{any} feasible statistical test.
More formally, any sufficiently long initial segment of an algorithmically random infinite sequence is incompressible (except by a fixed constant) or, equivalently, the sequence does not belong to the infinite intersection of any Martin-L\"of test~\cite{Calude2002main,Downey2019}.
As a unidimensional example in the context of sequences, algorithmic complexity theorists very soon realised that an object such as $ 123456789101112 \dots $ could be very misleading in terms of complexity. Note that this sequence in fact defines the Champernowne constant  $C_{ 10 } = 0. 123456789101112$, a complexity-deceiving phenomenon from the Borel normal numbers~\cite{zenil_kiani_2017} that is generated by one of the most modular forms of a function type, recursion and iteration of a successor-type function $ f( x_0 , x_{ i } ) = x_i + x_0 = x_{ i + 1 } $ for $ x_0 = 1 $. 
The ZK graph\cite{zenil_kiani_2017} which is constructed using the Champernowne constant as the degree sequence, was shown to be a near-maximal entropy graph with low algorithmic complexity~\cite{zenil_kiani_2017}. The reader is invited to note how such a mathematical concept motivated the construction of deceiving molecules in Section~\ref{sectionSupInfDeceivingmolecules}.
%% Beginning of Felipe's content addition.
%% About deceiving molecules

%% \todo{moved from the main text}
In~\cite{salientcronin}, the authors of Assembly Theory attempted to address what they thought were some concerns about an early preprint version of this manuscript. According to the authors, the statistical tools used were not appropriate, and the figures and comparisons above confound different types and sources of data with their application being unique in that it can be applied to data closer to what they think is the physical process, this is not correct. Regarding the statistical tools, even conceding (which we are not), that other statistical comparison methods are possible, it is not possible to manufacture desired results with one tool that with another one is found to be completely different. The authors seem to suggest that the greatest value of this paper is that some measures outperformed their Molecular Assembly (MA) index, and not that every other statistical index including some introduced in the 60s, applied to any chemical input data (not only mass spectra), reproduces and replicates what their paper~\cite{marshall_murray_cronin_2017} introduced as its main contribution.

\section{On dictionary-based algorithms}\label{sectionSupInfComparingMA}
%\section{Computational aspects of the results}\label{sectionSupInfComputationalaspects}

The assembly method derived from the `Assembly Theory' proposed by the original authors~\cite{cronin} consists roughly in finding a pattern-matching generative grammar behind a string by traversing and counting the number of copies needed to generate its modular redundancies, decomposing it into the statistically smallest collection of components that reproduce it without loss of information by finding repetitions that reproduce the object from its compressed form. 

For purposes of illustration, let us take the example of ABRACADABRA, which the original authors have also used~\cite{cronin}. For molecular assembly (MA) to succeed it needs to have a discriminator and classifier able to characterise each repetition of $A$ and $N$ as the same, where $N$ is another character or some sub-unit of the structure with the same frequency as $A$ (e.g., a two-letter unit containing $A$, such as $AB$ or $RA$).  
%This may be relatively possible for small molecules, allowing for some margin of additive noise, but it is not possible with anything beyond certain simple molecules.  
%Even for simple molecules, as we have shown in Section~\ref{sectionSupInfComparingMA}, a collection of other measures performs comparably or better than the measure from Assembly Theory, and this measure is also suboptimal in the general case.
%In their ABRACADABRA example, any minor modular generator that does not produce a simple statistical pattern will be missed.  
%To put this in a biological context, let's take cell membrane, they will be all of different type and size, yet are highly modular.  
%This is because weak measures will only able to characterise very low level simplistic types of modules that were already captured by simplistic measures, and from which the field moved on in the context of the characterisation of life.
In the ABRACADABRA example, MA deconstructs the sequence into unique blocks of five possible characters by adding a new character in subsequent steps, such that the minimal number of steps, considering only the frequency of the largest repeated block size (ABRA) just as it is done for LZ compression algorithms. The repeated binary or tertiary recursive structures (i.e., blocks of 2 or 3 letters) within the sequence, such as AB, RA, or BRA, are ignored in MA's minimal path search as in LZ.

The proposed Molecular Assembly (MA)~\cite{marshall_murray_cronin_2017}, the assembly index, and the Assembly Theory in general~\cite{cronin} fall into the category of dictionary-based entropy encoding indexes and approaches and are indistinguishable from an implementation motivated by and based upon the principles of algorithmic complexity using LZ compression.
In Section~\ref{sectionSupInfComparingMA}, we have shown the behavioural similarity between statistical compression and MA results consistent with the theoretical findings~\cite{abrahao2024}. These popular statistical algorithms have been universally used for data compression and as computable estimations to algorithmic complexity~\cite{liandvitanyi} and logical depth~\cite{zenilld,soler} including for classification purposes of living systems~\cite{cilibrasi,zenilchem,zenilld} of which AT is a special and, mathematically speaking, weak estimation. 

\begin{figure}[ht!]
	\centerline{\includegraphics[scale=0.35]{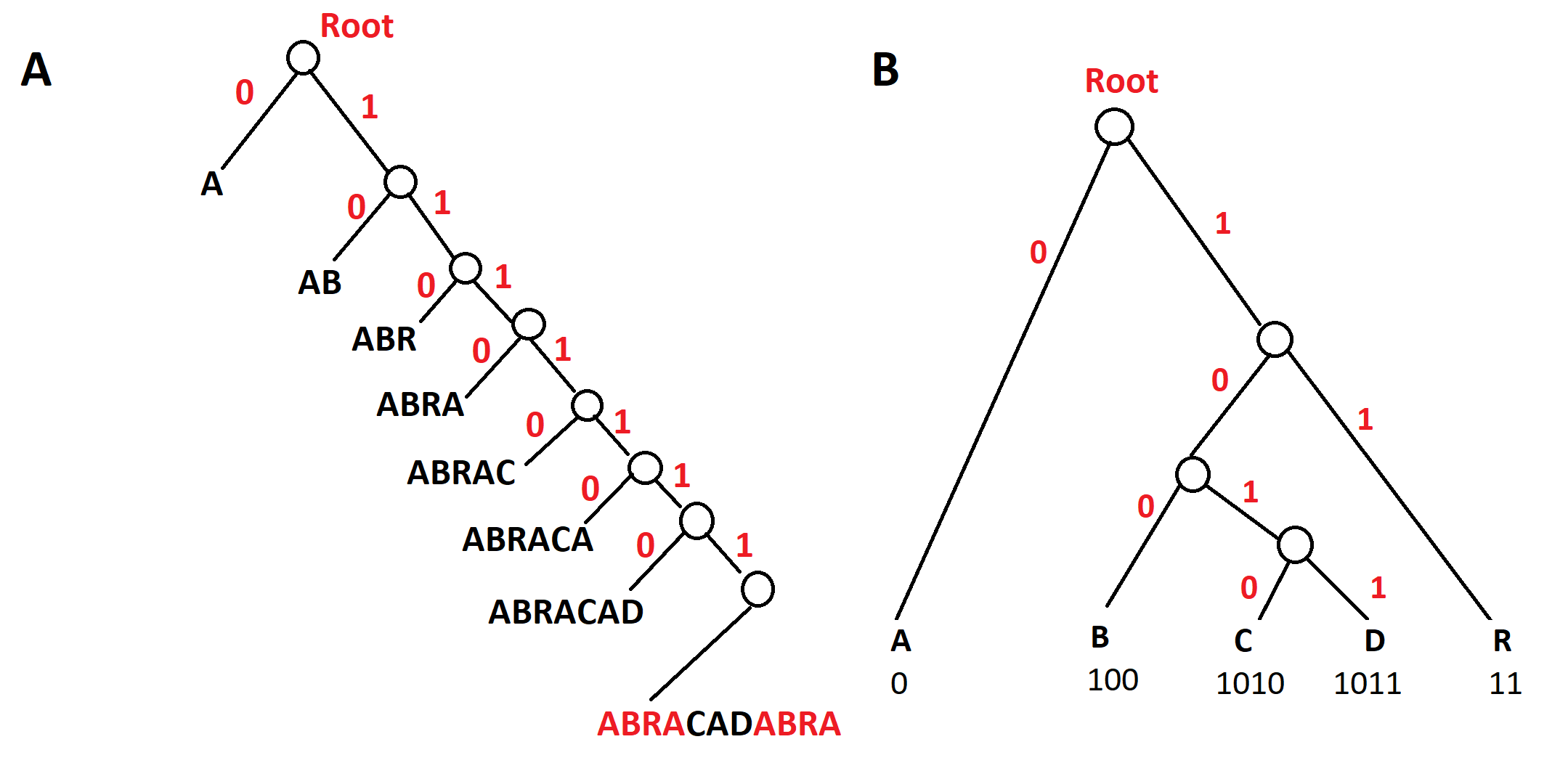}}
	\caption{\label{SupInffigureTreediagrams} ABRACADABRA tree diagrams for AT (A) and dynamic Huffman coding (B), both computable measures trivial to calculate. Huffman's was the first dictionary-based coding algorithm and is an optimal coding method able to characterise every statistical redundancy, including modularity, independent of such copy data representation~\cite{abrahao2024}.The (molecular) assembly index has been proven to be equivalent to LZ77/LZ78~\cite{abrahao2024}. In this example, Huffman's (which is also a sequential lossless compression algorithm that traverses strings from left to right) collapses the compression tree into a 4-level tree, while MA's is a 7-level tree. No natural evidence indicates that the assembly index (or MA) corresponds better to how nature works. However, the assembly index is identical to LZ compression~\cite{abrahao2024}. In both cases, the resulting tree of this word problem characterises the same token and is able to reconstruct it in full, without any loss of information, by exploiting redundancy (identical copies) producing a set of possible cause-and-effect chains for which no empirical evidence exists in support of MA. Both LZ and Huffman, just as MA, converge to the same Shannon Entropy rate and can be used to guide a search in chemical space.}
\end{figure}
%Thus, the Huffman tree is shorter and hence provides a shorter description of the information system (molecular biosignature). 
%Although algorithms characterise the number or set of unique blocks (i.e., characters or sets of unique symbols) in which the sequence or molecular structure can be decomposed to, 
Figure~\ref{SupInffigureTreediagrams} shows an illustration of the standard operation of Huffman coding in a typical example, compared to the principle advanced by the AT authors~\cite{cronin}. Proposed in the 50s, the Huffman coding exploits block redundancy by parsing objects, counting block recurrence and was one of the first, if not the first, dictionary-based coding algorithm~\cite{huffman_1952}. 

As shown in Fig.~\ref{SupInffigureTreediagrams} featuring the ABRACADABRA example, to the left (1A), we see the reconstruction of the sequence from a root node by the method proposed by AT, in general, and in this particular molecular application, when represented as a tree search diagram following binary branching rules. A bifurcation to the right denoted by 1 indicates a new assembly step, whereas a bifurcation to the left indicated by 0 from a node represents a fixed structure (block). The MA algorithm requires seven assembly steps to derive the sequence of interest. However, as shown to the left in Fig.~\ref{SupInffigureTreediagrams}, the Huffman coding tree optimises the sequence reconstruction by principles of recursivity in its search compression, as evidenced by the nested bifurcations. 

Unlike algorithms like Huffman, MA lacks bifurcations in the assembly search and instead considers a combinatorial search space with a linear sequence progression that cannot be justified by causal chain progression because it may not have anything to do with how an object may have been assembled through many parallel processes. The authors, therefore, conflate causation with plausible assembly ordering. 

Hence, it shirks the quantification of emergent hierarchical or nested structures (i.e., modularity optimisation) and intermediate structures within the sequence decomposition/compression. In contrast to MA, the recursiveness observed in complex molecules and biosignatures is detected by RLE and Huffman coding, and it does this in the most optimal way by providing the shortest tree algorithm (what the call `assembly pathways') needed. 

%We suggest that the reason behind this is that modularity, `nestedness', or recursion are inherent to the binary tree search framework of the Huffman coding algorithm that implements `counting copies'. 
%Thus, in this analysis MA was found to produce an expanded and suboptimal version of the Huffman coding tree for the same purpose, searching for the minimal path length (steps) to obtain a structure or sequence while considering only the frequency of the largest block size (e.g. ABRA) in the optimisation of the search. 

The results show that MA performs as any other coding/compression scheme because it is a compression scheme (even if they may have not intended to be so). The predictive power of MA, even if not unique or different, is therefore due to its information-theoretic properties based upon statistical compression of repeated patterns.

%On the other hand, the Huffman coding shows the emergence of all unique blocks, including recursive sub-structures (and their respective frequencies), in the shortest number of steps as a tree diagram with its shortest description, which is what AT and molecular assembly seemed intended to originally capture and what the authors designate as 'counting the number of copies'~\cite{cronin}.

%Hence, Huffman, though a very simple compression algorithm, is potentially a better coding scheme instantiating what AT intended but failed to implement, namely, to count the minimum of expected steps~\cite{Cover2005}, capturing the chained redundancy in an object.

%Huffman shows the emergence of these 2- or 3-letter blocks, and yet it manages to represent all this information in a minimal tree diagram. Unlike MA, Huffman coding detects nestedness (recursive blocks) within the molecular structure/sequence of interest. Huffman preserves both modularity and frequency of blocks within the shortest tree, whereas MA only looks at the minimal linear sequence optimised for only the largest repeated block size.

%\newpage
%\section{Supplementary Note 2}

\section{Methods}\label{sectionSupInf-Methods}

The list of MA values for all mass spectral signatures is available in~\cite{cronin}. 
Various complexity measures were used to classify living versus non-living molecules from the chemical space data in a four-category scheme: natural compounds, metabolites, pharmaceuticals, and industrial compounds, where the natural compounds include the amino acids, consistent with the classifications of Figure 2 in~\cite{cronin}. 
Further, the mass spectrometry (MS) data of the mixtures (biological extracts and non-living molecules) were categorized into three categories: abiotic, dead, and biological, as consistent with Figure 4 from~\cite{cronin}. The results were subjected to statistical analyses such as the Kolmogorov-Smirnov test, one sample t-tests, and Pearson correlation analysis using GraphPad Prism v. 8.4.3. 

Mass spectrometry (MS) data of the extracts and the various molecules used to construct the chemical space for validating MA theory, were analyzed using various complexity measures, including the 1D-string and 2D-matrix Block Decomposition Method (BDM)~\cite{bdm2,bdm,Zenil2020cnat}, Shannon’s entropy, and compression algorithms, including Lempel-Ziv-Welch (LZW), All data were first binarised using the online text-to-binary converter with ASCII / UTF-8 character encoding. Run Length Encoding (RLE), Huffman coding, and gzip. The InChI strings of the 99 molecules from (MW \textit{vs}$.$ MS data) of Fig.~2B, and the 114 molecules from Figure 3 (MS data standard curve) in~\cite{cronin} were binarised and analysed using the OACC (Online Algorithmic Complexity Calculator) app in R, which computed the 1D-BDM (block size of 2, alphabet size of 2, block overlap of zero) and Shannon Entropy scores. 
%In Figure 3, we assessed the molecules in~\cite{marshall_murray_cronin_2017}. 
The LZW compression lengths were computed with an online LZW calculator using UTF-8 encoding for the 1D strings. Likewise, RLE and Huffman coding compression lengths were obtained using online calculators as additional lossless compression measures to assess the MS bio-signatures. The RLE calculator was set to character then count settings, while the Huffman coding calculator output was set to compression ratio. As for the Figure~\ref{figureBiologicalComplexityplot}, biological extracts (mixtures) analysis, we used the mass spectra peak matrices of the mixtures (MS2 peaks \textit{vs}$.$ the total number of peaks) for the above-discussed method/analysis, post-binarisation above the threshold.

For the unpaired (two-samples/independent measures) t-test with Welch’s correction, at a degree of freedom (df) of 100, a critical t-value of 3.390 is expected for a two-tail P-value of 0.001 (i.e., 99.9\% confidence). The t-value closest to 3.39 was found for the 1D-BDM and 2D-BDM~\cite{bdm,2d}, with a t-value of 6.410 and 6.561, respectively (P $<$ 0.0001), both within the critical region of statistical significance. All complexity measures obtained a non-parametric Kolmogorov-Smirnov test value of P $<$ 0.0001; the Kolmogorov-Smirnov distance D was smallest for the 1D-BDM and 2D-BDM, with both returning a value of 0.707.

Through these statistical assessments, the 1D-BDM and 2D-BDM at a binary conversion threshold of 3 were found to be robust discriminants of molecular complexity in classifying living \textit{vs}$.$ non-living molecules. The result is shown in Figure~\ref{figureLivingversusNonliving}.

To perform the 2D-BDM on the MS signatures (molecules), the structural distance matrix was extracted from the 2D-molecular structure SDF files for each molecule using the PubChem database. Binary conversion was performed on the matrices after flattening in R at five different conversion thresholds (i.e., -1, 0, 1, 3 and 5). The matrices were flattened by taking rows of values for the distance matrices, or pairs of values for the MS2 matrices (for Fig.~\ref{figureBiologicalComplexityplot}) and ordering them by rows onto the Rapidtables ASCII/Unicode text string conversion calculator, and removing the spaces. The matrices can be simply pasted as a text, and converted to binary strings using the ASCII/UTF-8 character encoding and Space Output delimiter string settings. The binarised molecular distance matrices were processed by the PyBDM code to obtain the 2D-BDM scores for each molecule. Distance matrices at a binary conversion threshold of 3 were found to be optimal in differential analysis of the MA chemical space signatures and MS signatures into life \textit{vs}$.$ non-life categories. The matrices at a conversion threshold of 3 were used to compute the 2D-Huffman code and 2D-RLE compression lengths. 

%The theoretical background, definitions, theorems and proofs to which the above sections are referencing can be found in (See Supplementary Information). 

\subsection{Description of Algorithmic Complexity Measures}

The paper by Marshall et al.~\cite{marshall_murray_cronin_2017} introduced the molecular assembly (MA) algorithm as a new approach to assess the complexity of biochemical interactomes and distinguish biosignatures of life from abiotic chemistry. The hypothesis is that the molecular assembly (MA) algorithm models the assembly process of biomolecules.

To evaluate the MA algorithm's utility for analyzing complex biochemical systems, we benchmarked it against established methods from algorithmic information theory that quantify algorithmic complexity. Algorithmic information theory, rooted in computational complexity theory and Kolmogorov complexity, provides a rigorous mathematical framework to assess the information content of objects independent of the specific encoding. We applied several algorithmic information theory techniques to the same datasets analyzed in~\cite{marshall_murray_cronin_2017}, including molecular distance matrices and mass spectrometry data of chemical interactomes. Our toolkit consisted of:

1D-BDM: The block decomposition method (BDM) segments data vectors into variable blocks and calculates the algorithmic complexity from the entropy of block sizes~\cite{bdm}. 1D-BDM analyzes one-dimensional data such as DNA sequences. 
2D-BDM: The two-dimensional BDM acts on distance matrices and other 2D datasets. It divides matrices into rectangular blocks and computes complexity from block entropies.

LZW: The Lempel-Ziv-Welch (LZW) algorithm evaluates complexity by the number of steps needed to compress data (Ziv and Lempel, 1978). It builds a dictionary of sequences and measures complexity by dictionary size. 
Shannon Entropy: As a measure of uncertainty, or randomness, it quantifies information content based on symbol frequencies in data. Lower entropy indicates simpler patterns.

1D-RLE: Run-length encoding (RLE)~\cite{Robinson1967ResultsPrototypeTelevision} compresses data by counting repetitions of symbols. 1D-RLE counts runs in 1D sequences.
1D-Huffman: Huffman coding~\cite{huffman_1952} compresses data by assigning shorter codes to more frequent symbols. 1D-Huffman operates on 1D sequences.

2D-RLE and 2D-Huffman: The 2D versions apply RLE and Huffman coding to 2D matrices by encoding runs or frequencies of matrix rows/columns.

%This algorithmic information toolkit provides compression-based and statistical measures of algorithmic complexity grounded in information theory. They quantify the information required to losslessly reconstruct data, revealing deep patterns in the informational entropy of systems (Zenil et al., 2019). Prior work has successfully applied these techniques to characterise complex biosignatures in omics data and biological sequences (Soler-Toscano et al., 2014).

Our results reveal fundamental limitations in using path lengths on assembly trees as a proxy for complexity. MA is most correlated to 1D-Huffman coding, which encodes more probable symbols with shorter codewords. Both exploit the frequency/probability distribution of components. However, this is just one facet of algorithmic complexity. Our toolkit provides a more complete view, with LZW, BDM, and 2D techniques assessing additional dimensions such as long-range sequence structure and 2D constraints. 
In summary, this paper benchmarks the novel MA algorithm against established algorithmic information measures for quantifying complexity in biochemical interactomes. 

Our results reveal that MA has salient limitations as a standalone measure for biosignature complexity. The diverse measure toolkit provides a more robust perspective on assembly processes and algorithmic complexity in chemical systems. Our work elucidates the connections between MA and lossless compression, while demonstrating the power of algorithmic information measures that take into account regularities beyond trivial statistical redundancies. 
%Further information on the algorithmic complexity measures can be found in our Supplementary Information.

\subsection{PyBDM Code for CTM and BDM}

The Coding Theorem (CTM) and Block Decomposition Methods (BDM) are resource-bounded computable methods~\cite{soler,bdm,review} that attempt to approximate semi-computable measures that are a generalisation of statistical measures more powerful than the methods proposed in ``Assembly Theory'' as they combine global calculations of classical entropy with local estimations of algorithmic information content.

\begin{algorithm}[ht!]
\caption{\label{SupInfalgorithmPyBDM}Python implementation of 2D-Block Decomposition Method (PyBDM)}\label{SupInfPyBDM}
import numpy as np\\
from pybdm import BDM\\
import pandas as pd\\
X = pd.read\_csv(r'file directory',dtype=int)\\
bdm = BDM(ndim=2)\\
Z=X.to\_numpy()\\
bdm.bdm(Z) 
\end{algorithm}

%\section{MS Classification Trends in Molecular Complexity Measures by Molecular Weight identifies 1D-BDM and 2D-BDM as optimal discriminants of life \textit{vs}$.$ non-life}
%
%The comparison of complexity measures in Fig.~\ref{SupInffigureComplexityclassification} across the four categories of MS molecules is shown in relation to increasing molecular weight to better visualize the trends across living and non-living bio-signatures (Fig.~\ref{figureLivingversusNonliving}).
%\todo[inline]{Could this figure be embedded into the main manuscript?}
%\begin{figure}
%\centerline{\includegraphics[scale=0.35]{Figure_S1_Complexity_Trends_by_MW.png}}
%\caption{\label{SupInffigureComplexityclassification}MW Trends in Complexity Measures amidst the four categories of MS dataset}
%\end{figure}

\section{Expected false positives from complexity-deceiving molecules with arbitrarily high statistical significance}

As we move beyond the realm of pure stochastic processes, complexity distortions become more problematic \cite{zenil_kiani_2017,Abrahao2021publishednat,Abrahao2021darxiv,Abrahao2020cAIDistortionsCN}.
As demonstrated in Theorem~\ref{SupInfthmDeceivingmolecule} and Corollary~\ref{SupInfthmDeceivingmoleculeinauniversalspace}, there are (sufficiently large) deceiving molecules the complexities of whose respective generative processes arbitrarily diverge from the assembly index that the assembly pathway method assigns to them.
We demonstrate in Section~\ref{sectionSupInfDeceivingmolecules} that there are (sufficiently large) deceiving molecules the complexities of whose respective generative processes arbitrarily diverge from the assembly index that the assembly pathway method assigns to them.
By a generative process~\cite{Abrahao2021bpublished} we mean any process that can be implemented, computed, or emulated by another equivalent (or identical) process so that it generates the pathway assembly and its object.
In accordance with the assumptions and rationale in \cite{cronin,marshall_murray_cronin_2017,Marshall2019}, those generative processes are exactly those composed of the assembling processes allowed (i.e., deemed physically possible by assembly theory's chosen method) so that they in the end result in the constitution of molecules.

Notice that our results in Section~\ref{sectionSupInfDeceivingmolecules} hold for any of such processes that belong to (abstract or physically implemented) computational classes much weaker than that of Turing machines, e.g. resource-bounded Turing machine \cite{review,Allender2011,soler,Calude2002main,Burgin2009,liandvitanyi} or simple forms of finite automata \cite{Wolfram2002,Hutchison2006}.
Nevertheless, if the class of generative processes is constituted by processes that are also capable of universal computation, then Corollary~\ref{SupInfthmDeceivingmoleculeinauniversalspace} shows the complexity distortions (and, therefore, also the deceiving phenomena we explain below) can be equally bad or even worse.
Therefore, AT fails to capture, both in theory (when resources can be unbounded) and in practice (when resources are limited), the minimality that is necessary for a complexity measure that may be claimed to be unambiguous and observer-independent.
% Additionally, MA in general fails to avoid false positives in the specific sense that it may not be able to distinguish a ``complex'' object that is in fact the outcome of randomly generated (resource-bounded) generative processes.

While the calculation of MA may be prone to false negatives---due to partial fragmentation in energy collision analysis and the restriction to counting only valence rules in molecule synthesis (ignoring other chemical conditions)---this does not pose a challenge to the central claim made in~\cite{cronin}. 
Instead, MA aims at avoiding underestimation of the amount of molecules that result from random or abiotic processes, hence aiming at avoiding false positives. 
Thus, by directly tackling their central claims and this motivation, one demonstrates in Theorem~\ref{SupInfthmDeceivingmolecule} that MA in general fails to avoid false positives in the specific sense that it may not be able to distinguish a ``complex'' object that is in fact the outcome of randomly generated (resource-bounded) generative processes.

Under the same assumptions as in~\cite{cronin,Marshall2019}, we construct in Theorem~\ref{SupInfthmDeceivingmolecule} a deceiving molecule that has a much larger MA value in comparison to the minimal information sufficient for a randomly generated generative process to single-handedly construct this molecule.
Whatever arbitrarily chosen method is used to calculate the statistical significance level, the MA of this molecule is large enough to make the expected frequency of occurrence (estimated via the arbitrarily chosen AT) diverge from the actual probability.
In this case, AT would consider such a molecule ``biotic'', resulting from extrinsic factors that increase biases toward certain pathways or that constrain the range of possible joining operations, although its sole underlying generative process in fact results from fair-coin-toss random events.
This proven existence of false positives due to such a deceiving phenomenon is corroborated by our empirical findings, which show that MA displays a behaviour that is both structurally and empirically similar to traditional statistical compression methods.
Indeed, the latter methods are already known to present distorted values~\cite{zenil_kiani_2017}, performing worse than more recent algorithmic-based methods~\cite{review}. Thus, they are prone to overestimating complexity, and consequently to presenting false positives for ``high''-complexity objects which are in fact simple.

The key rationale behind this result is the computable nature of MA, so that given the set of biases and the joining operations allowed by the model, objects with much higher MA can be constructed by much simpler (and, therefore, more probable) computable generative processes which can in turn be randomly generated.
The more computationally cheap and tractable the AT's method is, the lower the complexity and the more limited the resources in which the deceiving phenomenon is expected to take place.
Thus, in this context, MA (or any computable `assembly' measure of this basic statistical type) will underestimate the frequency of occurrence of objects with high MA that in fact were constructed by much simpler randomly generated processes.
This means that MA would misidentify molecules as byproducts (or constituents) of livings systems that resulted from evolutionary processes, while in fact these molecules might have been byproducts of single-handed computable systems that were randomly generated by a fair coin toss, and as such are not the result of an evolutionary process of optimisation over time.

Nevertheless, note that it is true that there are computable (lossless) encodings of a source, such as Huffman coding, that are proven to be optimal on average, but only if one knows a priori that the underlying processes generating the objects are purely stochastic (in particular, when one knows beforehand that the conditions of the source coding theorem are satisfied~\cite{Cover2005}).
In this case, one can show that the minimum expected size of the encoded object converges to its expected algorithmic complexity \cite{Calude2002main,Cover2005}.
However, pure stochasticity is too strong an assumption, or does not realistically represent the generative processes of molecules.

On the one hand, living systems are complex systems consisting of multiscale, multi-nested processes that are unlikely to be reducible to simplistic and intrinsic statistical properties such as those suggested by AT.
One cannot conceive a measure that only looks at the internal structure of an agent isolated from its environment and how it interacts with its external medium to determine its (non)living nature.
Especially in the context of complex systems like living organisms, organic molecules may be the byproduct of intricate combinations or intertwinements of both deterministic/computable and stochastic processes that govern the behaviour of the entire organism and its environmental surroundings~\cite{Walker2012nat,bdm}.

On the other hand, in case a sufficiently complex (e.g., at the level of those we demonstrate in Section~\ref{sectionSupInfDeceivingmolecules}) environmental catalytic condition plays the role of the above extrinsinc factor (which increases the bias toward the construction of a more complex molecule), this level of complexity would be completely missed by the capabilities of simplistic measures such as MA, thereby rendering it as prone to false positives.
More careful and deeper arguments regarding simplicity, recursivity, and the emergence of modularity in life have been advanced and are better grounded in a theoretical and methodological framework advanced in~\cite{santiago}, where it was shown that exploiting first principles of computability and complexity theories, modular properties in evolutionary systems may be explained. 

As shown by Corollary~\ref{SupInfthmDeceivingmoleculeinauniversalspace}, the deceiving phenomena can be equally bad or even worse in case the molecules are byproducts of complex systems that are somehow capable of universal computation.
For example, in the case of advanced civilisations that are capable of artificially constructing living beings by computable processes, simplistic complexity measures such as MA can be intentionally misled with respect to what actually should be measured;
or in the case of extrinsic environmental catalytic processes whose chemical dynamics' complexity are comparable to those of some randomly generated finite automata \cite{Wolfram2002}.
%Unless one accepts to be prone to the deceiving phenomenon we demonstrate in this article, there is always room for improvement, as the best one can do is to develop new methods that can better harvest the most computational resources available to the moment.
%Thus, contrary to the authors' claim in \cite{croninmain,Marshall2019main} that computability and tractability would be an advantage, the more tractable and simpler a computable complexity measure, the easier (in terms of computational resources) it is to deceive this complexity measure.

%In this way, assembly pathway theory does not give an optimal complexity measure in the general case, as it may return a misleading value that would classify a low complexity molecule as being extrinsically constructed by a much complex agent, while in fact this extrinsic agent is much simpler and randomly generated.
Our results suggest the presence in more realistic resource-limited scenarios of a property known to occur in theoretical complex systems science \cite{Abrahao2021bpublished}:
in the context of generative processes that are not purely stochastic, and are capable of displaying complexity or computation capabilities at the same level of those of the observers, there is no such thing as a generally optimal complexity measure that cannot be improved upon, since computable complexity measures are dependent on the observer (or the chosen formal theory).
For example, without the necessary conditions being satisfied by the underlying stochastic process, one cannot generally guarantee such a convergence between the expected size of the encoded object and the expected algorithmic complexity that is assured by the source coding theorem.

\subsection{Mathematical framework and assumptions}\label{sectionSupInfDeceivingmolecules}

% The main idea to achieve the following theoretical results is to construct a randomly generated program that receives a formal theory (which contains all the computable procedures and statistical criteria in assembly theory) as input.
% Then, it searches for a molecule (or object in an assembly space) with MA sufficiently high so as to make the pathway probability of spontaneous formation be sufficiently lower than the very own deceiving program's algorithmic probability, so that the divergence between these two probability distributions become statistically significant according to the chosen statistical method and significance level.

In order to demonstrate the presence of false positives in such realistic scenarios where both complexity and resources are limited (and, therefore, every process and measure is computable), one should account for the cases in which only a subclass of possible computable processes is allowed to perform the assembly rules (e.g., those that are allowed by the currently known law of physics in the case of molecules) in order to construct molecules.
In this case, not every type of computable function may represent what is an effective or feasible  process that constructs a molecule.
Thus, in some cases the range of generative processes that can give rise to (or construct) a molecule may not comprise all possible computable functions.
With this purpose, we employ a variation of the traditional algorithmic complexity and algorithmic probability studied in Algorithmic Information Theory (AIT).
For this reason, we will employ a suboptimal form of the algorithmic complexity that is defined on non-universal programming languages (i.e., subrecursive classes).

Nevertheless, in an ideal (resource-\emph{unbounded}) case in which the whole algorithm space indeed constitutes the set of all possible generative processes for constructing the assembly space (e.g., when chemical processes are able to achieve the capability of effecting universal computation in the real world \cite{Walker2012nat,Wolfram2002}), we also show in Corollary~\ref{SupInfthmDeceivingmoleculeinauniversalspace} that the deceiving phenomenon hold in the same way (or can even be worse).

A deceiving phenomenon akin to the one employed in Theorem~\ref{SupInfthmDeceivingmolecule} can be found in \cite{Abrahao2021darxiv} based upon the same AIT principles in \cite{zenil_kiani_2017},
where sufficiently large datasets were constructed so that they deceive statistical machine learning methods into being able to find an optimal solution that in any event is considered global by the learning method of interest, although this optimal solution is in fact a simpler local optimum from which the more complex actual global optimum is unpredictable and diverges.

%Thus, contrary to the authors' claim in \cite{cronin,Marshall2019} that computability and tractability would be an advantage, the more tractable and simpler a computable complexity measure, the easier (in terms of computational resources) it is to deceive this complexity measure.
%Indeed, this applies not only to MA, but also to any computable measure for which the statistical significance of the sample can be computably tested (see \cite{SupplInfo} and \cite{Abrahao2021darxiv}).
This phenomenon is also related to the optimality of the algorithmic complexity as an information content measure that takes into account the entire discrete space of computable measures \cite{Downey2010,Chaitin2004}, or the maximality of the algorithmic probability as a probability semimeasure on the infinite discrete space of computably constructible objects, as demonstrated by the algorithmic coding theorem \cite{liandvitanyi,Downey2010,Calude2002main}.

However, unlike in these previous cases, our proof is based on finding a deceiver algorithm that constructs an object with sufficiently high value of assembly index such that its expected frequency of occurrence is much lower than the algorithmic probability of the deceiver itself, and in this way passing the test of any statistical significance level the arbitrarily chosen formal theory may propose.

In order to achieve our results, we base our theorems on mathematical conditions that are consistent with the assumptions and results in \cite{cronin,Marshall2019}.
The first \emph{assumption} that we specify with the purpose of studying a worst-case scenario is that the assembly space should be large enough so as to include those molecules (or objects) with sufficiently large MA (along with its associated sufficiently low pathway probability of spontaneous formation) relative to the algorithmic complexity of the deceiving program.
For the sake of simplicity, we assume that the nested family $ \mathcal{ S } $ of all possible finite assembly spaces from the same basis (i.e., the root vertex that represents the set of all basic building blocks) is infinite computably enumerable.
However, an alternative proof can be achieved just with the former---and more general---assumption that the assembly space may be finite but only needs to be sufficiently large in comparison to the deceiving program.
Indeed, our assumption is in consonance with the authors' motivation (and/or assumption) that 
``biochemical systems appear to be
able to generate almost infinite complexity because they have
information decoding and encoding processes that drive networks
of complex reactions to impose the numerous, highly specific
constraints needed to ensure reliable synthesis'' \cite{cronin}.

Closely related to the first assumption, we also \emph{assume} that there always are molecules with arbitrarily low path probabilities,
which follows from the notion that, as infinitesimal as it might be, there is always a chance of randomly combining elements from an unlikely (but possible) sequence of events so as to give rise to a certain complex molecule.

Thirdly, in accordance with the arguments in \cite{cronin,marshall_murray_cronin_2017,Marshall2019} that the computability and feasibility of their methods is an actual advantage in comparison with other complexity measures, here we likewise adopt these same \emph{assumptions} so that the following are computable procedures:
\begin{itemize}
	
	\item deciding whether or not a finite assembly space (or subspace) is well formed according to the joining operation rules that are allowed to happen;\footnote{ For example, one can employ the same criteria and allow the same rules established in \cite{cronin}.}
	
	\item calculating MA of a finite molecule (i.e., a finite object) in a well-formed assembly space (or subspace);\footnote{ For example, as defined in \cite[Definition 19]{Marshall2019}. }
	
	\item calculating the chosen approximation of MA (e.g., the split-branch version) of a finite molecule in a well-formed assembly space (or subspace);\footnote{ For example, as in \cite[SI]{cronin} and \cite[Section 4.2]{Marshall2019}.}
	
	\item calculating an upper bound for the pathway probability of spontaneous formation of a molecule in the denumerable nested family of possible finite assembly spaces;\footnote{ For example, this can be done by employing the methods developed in \cite[SI]{Marshall2019} and \cite{cronin}}
	
	\item calculating the significance level for a frequency of occurrence of a molecule in a sample so that this empirical probability distribution (i.e., the type of the sample) diverges from the pathway probability distribution of spontaneous formation of the molecule;\footnote{ For example, by using a maximum-likelihood method or using the probability that a sample occurs with KL divergence larger than $\epsilon$ \cite{Cover2005,Goodfellow2016}. }

	%	\item deciding whether or not a program represents the total recursive function that computes a generative process that gives rise to a molecule;\footnote{ For example, by employing the current knowledge of physics to decide whether or not the program encodes a real-world process that can be emulated by a Turing machine.}
	
\end{itemize} 

%In addition, we \emph{assume} that the set of possible underlying generative processes (or models) of the assembly spaces is computably enumerable, and each element in this set is equivalent to a computable system.
%This is because, given the current knowledge of physics, chemistry, and/or biology encoded into some formal theory, one should for example be able to enumerate the constructible or feasible processes that can possibly generate assembly spaces.

% In this manner, one can now demonstrate the following theorems.

\subsection{Definitions and notation}

Besides the notation from \cite{Marshall2019} for assembly theory, we also employ the usual notation for Turing machines and algorithmic complexity.

Respectively, as in \cite[Definition 11]{Marshall2019} and \cite[Definition 15]{Marshall2019}, let $ \left( \Gamma , \phi \right) $ denote either an \emph{assembly space} or \emph{assembly subspace}.
%Note that $ \Gamma $ is an acyclic multigraph and $ \, \phi \colon E( \Gamma ) \to V( \Gamma ) $ is an edge-labeling map.
From \cite[Definition 19]{Marshall2019}, we have that $ c_\Gamma\left( x \right) $ denotes the \emph{assembly index} of the object $ x $ in the assembly space $ \Gamma $.

Note that assembly spaces are finite.
So, from our assumptions, we need to define a pathway assembly that can deal with arbitrarily large objects. 
To this end, let $ \mathcal{ S } = \left( \mathbf{ \Gamma } , \mathbf{ \Phi } , \mathcal{ F }   \right) $ be an \emph{infinite assembly space}, where every assembly space $ \Gamma \in \mathbf{ \Gamma } $ is finite, $ \mathbf{ \Phi } $ is the set of the correspondent edge-labeling maps $ \phi_\Gamma $ of each $ \Gamma $, and $ \mathcal{ F } = \left( f_1 , \dots , f_n , \dots \right) $ is the infinite sequence of embeddings \cite{Hodges1993} (in which each embedding is also an \emph{assembly map} as in \cite[Definition 17]{Marshall2019}) that ends up generating $ \mathcal{ S } $.
That is, each $ \, f_i \colon \left\{ \Gamma_i \right\} \subseteq \mathbf{ \Gamma } \to \left\{ \Gamma_{ i + 1 } \right\} \subseteq \mathbf{ \Gamma } \, $ is a particular type of assembly map that embeds a single assembly subspace into a larger assembly subspace so that the resulting sequence of nested assembly subspaces defines a total order $ \preceq_{ \mathcal{ S } } $, where 
\[ 
\left( \Gamma_i , \phi_{ \Gamma_i } \right) \preceq_{ \mathcal{ S } } \left( \Gamma_{ i + 1 } , \phi_{ \Gamma_{ i + 1 } } \right)  \text{ \textit{iff} }   f_i\left( \Gamma_i  \right) = \Gamma_{ i + 1 } 
\text{ .}
\]

Let $ \gamma = z \dots y $ denote an arbitrary path from $ z \in B_\mathcal{ S } $ to some $ y \in V\left( \mathcal{ S } \right) $ in $ \mathcal{ S } $, where $ B_\mathcal{ S } $ is the basis (i.e., the finite set of basic building blocks) of $ \mathcal{ S } $ and $ V\left( \mathcal{ S } \right) $ is the set of vertices of $ \mathcal{ S } $.
Let $ \gamma_x $ denote a rooted path from some $ z \in B_\mathcal{ S } $ to the object $ x \in V\left( \mathcal{ S } \right) $.

Let $ { \Gamma^* }_{ x } $ denote a minimum rooted assembly subspace of $ \Gamma $ from which the assembly index $ c_\Gamma\left( x \right) $ calculates the augmented cardinality and that its longest rooted paths $ \gamma_x $ ends in the arbitrary object $ x \in V\left( \Gamma \right) $ as in \cite[Definition 19]{Marshall2019}.

As usual, let $ \mathbf{U} $ be a universal Turing machine on a universal programming language $ \mathbf{L} $.
Let $ \mathbf{U}(x) $ denote the output of the universal Turing machine $\mathbf{U}$ when $  x \in \mathbf{L} $ is given as input in its tape.
Let $ \left< \, \cdot \, , \, \cdot \, \right> $ denote an arbitrary recursive bijective pairing function \cite{Downey2010,liandvitanyi} so that the bit string $ \left<  x ,  y  \right> $ encodes the pair $ \left( x , y \right) $, where $ x , y \in \mathbb{N} $.
Note that this notation can be recursively extended to $ \left< \cdot \, , \, \dots \, , \, \cdot   \right> $ in order to represent the encoding of $n$-tuples.
%The Big-\textbf{O} notation $f(x)=\mathbf{O}( g(x) )$ denotes the usual \emph{weak} asymptotic dominance when function $f$ is asymptotically upper bounded by function $g$.

We have that the (prefix) \emph{algorithmic complexity}, denoted by
$ \mathbf{K}\left( x \right) $, is the length of the shortest prefix-free (or self-delimiting) program $ x^* \in \mathbf{L} $ that outputs the encoded object $ x $ in a universal prefix Turing machine $ \mathbf{U} $, i.e., $ \mathbf{U}\left( x^* \right) = x $ and the length $ \left| x^* \right| = \mathbf{K}\left( x \right) $ of program $ x^* $ is minimum.
%Other variants of algorithmic complexity in AIT include:
%the \emph{conditional} prefix algorithmic complexity of a binary string $ z $ given a binary string $ w $, denoted by $ \mathbf{K}( z \, \vert w ) $, which is the length of the shortest program $ z_w^*$ such that $ \mathbf{U}( \left< w , z_w^* \right> ) = z $;
%$ \mathbf{ I }_{ \mathbf{ K } }( w : z ) = \mathbf{K}(z) - \mathbf{K}( z \, | w ) $, which is the \textbf{K}-complexity of information in $ w $ about $ z $ and it quantifies the amount of irreducible information in $ w $ about $ z $; and
%the \emph{mutual algorithmic information}
%$  \mathbf{ I_A }( w \, ; z ) = \mathbf{K}(z) - \mathbf{K}( z \, | w^* ) $ between the arbitrary strings $w$ and $z$, which quantifies the amount of irreducible information in $ w $ about $ z $, and vice-versa.
In addition, the \emph{algorithmic coding theorem}  \cite{Downey2010,Chaitin2004,Calude2002main,liandvitanyi} guarantees that
\begin{equation}
	\mathbf{K}\left( x \right) = - \log\left( \sum\limits_{ \mathbf{U}\left( p \right) 
		= x } \frac{ 1 }{ 2^{ \left| p \right| } } \right) \pm \mathbf{O}( 1 )
	%= - \log\left( \mathbf{m}\left( x \right) \right) \pm \mathbf{O}( 1 )
	\text{ ,}
\end{equation}
where
%$ \mathbf{m}\left( \cdot \right) $ is a maximal computably enumerable semimeasure; 
$ \sum\limits_{ \mathbf{U}\left( p \right) 
	= x } 2^{ - \left| p \right| }  $
is the \emph{universal a priori probability} of $ x $, which gives the probability of randomly generating (by an i.i.d. stochastic process) a prefix-free (or self-delimiting) program that outputs $ x $. 
We also have it that $ 2^{ - \mathbf{K}\left( x \right) } $ is called the \emph{algorithmic probability} of $ x $, which therefore converges to the universal a priori probability (except of an object-independent constant).

If the language $ \mathbf{L}' $ is a proper subset of $ \mathbf{L} $ such that language $ \mathbf{L}' $ running on  $ \mathbf{U} $ is not able to decide every problem of Turing degree $ \mathbf{0} $, then we have that $ \mathbf{L}' $ is \emph{not} a universal programming language, 
and the machine $ \mathbf{U} $ defined upon language $ \mathbf{L}' $ is a \emph{Turing submachine} $ \mathbf{U}/f $ \cite{Abrahao2016nat}, where $ f $ is the partial function that computes the function $ \mathbf{U}\left( x \right) $ for $ x \in \mathbf{L}' $ as input.
In other words, a Turing submachine is a Turing machine that can receive inputs in its tape (and possibly simulate other machines), but it is not universal.
Weaker than Turing degree $ \mathbf{0} $, a submachine can only compute problems in a subrecursive class of problems \cite{Abrahao2016nat}.
Thus, note that resource-bounded machines or total Turing machines are particular cases of Turing submachines \cite{Abrahao2016nat}.

As a consequence of the above definitions, we define the (prefix) \emph{sub-algorithmic complexity}\footnote{ See also \cite{Abrahao2016nat} where this terminology is also employed.} $ \mathbf{K}_f\left( x \right) $ to be the length of the shortest prefix-free (or self-delimiting) program $ x^* \in \mathbf{L}' $ that outputs the encoded object $ x $ when run on the Turing submachine $ \mathbf{U}/f $ (i.e., $ \mathbf{U}/f\left( x^* \right) = x $ and $ \left| x^* \right|_f = \mathbf{K}_f\left( x \right) = \min \left\{ \left| w \right| \, \middle\vert \, \mathbf{U}\left( w \right) = x , \, w \in \mathbf{L}' \right\} $).\footnote{ If there is no program in $ \mathbf{L}' $ that can output an object $ x $, then one defines $ \mathbf{K}_f\left( x \right) = \infty $.}
Thus, note that resource-bounded variants of the algorithmic complexity \cite{Calude2002main,Burgin2009,liandvitanyi} are particular cases of sub-algorithmic complexity.
Analogously, we will have that the \emph{sub}-universal (a priori probability) distribution upon language $ \mathbf{L}' $ is defined on the \emph{sub}-universal a priori probability for each value $ x $, which are given by
\begin{equation}
	\sum\limits_{
		\begin{array}{c}
			\mathbf{U}\left( p \right) = x; \\
			p \in \mathbf{L}'
	\end{array}  } 
	C \, \frac{ 1 }{ 2^{ \left| p \right| } }
	\text{ ,} 
\end{equation}
where $ \mathbf{L}' \subseteq \mathbf{L} $ and $ C \geq 1 $ is a normalizing constant as in \cite[Definition 3.6, Section 3.2.1]{Abrahao2017publishednat} to ensure it is a probability measure and not a probability semimeasure. 
Note that, if $ \mathbf{L}' = \mathbf{L} $, then one obtains the usual universal distribution instead of its subrecursive version.

Following these notions, we can now define the submachine that compute the allowed (physical, chemical, and/or biological) \emph{generative processes} of an assembly space that assembly theory may arbitrarily choose.
Following the same rationale from \cite{cronin,marshall_murray_cronin_2017,Marshall2019}, those generative processes are exactly composed of the processes allowed by our current knowledge of physics so that they in the end result in the constitution of molecules.

As assumed to the above, and in accordance with the claims in \cite{cronin,marshall_murray_cronin_2017,Marshall2019}, notice that the computability and feasibility of assembly theory's methods directly implies the existence of such a submachine.
Given any set of (physical, chemical, and/or biological) rules for assembling molecules that assembly theory arbitrarily chooses to be based on and that is consistent with our current knowledge of physics, one has it that for any particular pathway assembly resulting in a molecule, there is a correspondent program running on a submachine $ \mathbf{U}_\Gamma $ for which one can apply the methods described in \cite{cronin,marshall_murray_cronin_2017,Marshall2019} to calculate the pathway probability and so on.
This existence is trivially guaranteed to hold as long as the calculation of the assembly index is computable given an assembly space, which is already an assumed condition in \cite{cronin,marshall_murray_cronin_2017,Marshall2019};
moreover, any upper bound for the computational resources necessary to compute the assembly index straightforwardly implies the existence of an upper bound for the computational resources necessary for a program to run on $ \mathbf{U}_\Gamma $ (i.e., submachine $ \mathbf{U}_\Gamma $ is a type of resource-bounded Turing machine).

For the sake of simplifying notation, $ \mathbf{U}_\Gamma $ denotes the Turing submachine $ \mathbf{U}/{ f }_\Gamma $.
In this case, the function $ f_\Gamma $ is the partial function that returns what $ \mathbf{U} $ can compute with some $ x \in \mathbf{L}_\Gamma $ as input, where $ \mathbf{L}_\Gamma \subseteq \mathbf{L} $ is a (non-universal) programming language such that every allowed generative process---i.e., every generative process that is deemed physically possible by assembly theory's chosen method, in case the objects are molecules---of an assembly space is bijectively computed (or emulated) by a corresponding $ \mathbf{U}\left( x \right) $.
In other words, for every allowed generative process that can assemble objects into building another object, there is a program $ x \in \mathbf{L}_\Gamma $ that computes (or emulates) this process.
Conversely, for every $ x \in \mathbf{L}_\Gamma $, one also has it that there is a corresponding generative process allowed by assembly theory, process which is computed (or emulated) by program $ x \in \mathbf{L}_\Gamma $.
Notice that in the forthcoming Theorem~\ref{SupInfthmDeceivingmolecule}, we will show that if a certain generative process is equivalent (in terms of complexity and computation capability) to a particular type of algorithm running on this resource-bounded submachine, then the deceiving phenomenon takes places.
Although our theoretical results in the next section prove the existence of false positives in the case this condition is met, our empirical results shown in Section~\ref{sectionSupInf-Results} not only corroborate this prediction but in fact suggest this theoretical finding (or some weaker version of it) may be happening in much lower computational classes.
This occurs because the assembly index either underperforms or does not outdistance other measures of a statistical nature, where distortions in complexity estimations are already well known to occur---which in particular is one of the reasons complexity science moved on from simplistic statistical measures and remains an evolving field of research.

In the case of infinite assembly spaces, one analogously defines language $ \mathbf{L}_\mathcal{ S } \subseteq \mathbf{L} $.
(In the special case in which the generative process of the assembly space $ \mathcal{ S } $ are capable of universal computation, then one has that $ \mathbf{L}_\mathcal{ S } = \mathbf{L} $ holds).
Also for the sake of simplicity, let $ \mathbf{K}_\Gamma $ denote the sub-algorithmic complexity $ \mathbf{K}_{ f_\Gamma } $.
That is, $ \mathbf{K}_\Gamma\left( x \right) $ gives the shortest program that can compute or emulate a generative process of the object $ x $ in the assembly space $ \Gamma $.
In the case of infinite assembly spaces, one analogously defines the sub-algorithmic complexity $ \mathbf{K}_\mathcal{ S } $ and the sub-universal a priori probability upon language $ \mathbf{L}_\mathcal{ S } $.

%When dealing with other kind of objects that are not strings, a mathematical object is said to be \emph{encoded} if it is univocally represented by structured data so that there is an algorithm which can always recover or extract the original object from the structured data.
%A trivial example is encoding a (directed) graph as an indexed list of the characteristic function of the edges in the form 
%\[ \left( \left( v_1 , v_2 \right) , z_1 \right) \cdots  \left( \left( v_i , v_j \right) , z_k \right) \cdots  \left( \left( v_{ n } , v_{ n - 1 } \right) , z_{ n^2 - n } \right) \text{ ,} \] 
%where $ n $ is the number of vertices, $ n^2 - n $ is the total number of possible (directed) edges, $ 1 \leq i \leq n $, $ 1 \leq j \leq n $, and  $ 1 \leq k \leq n^2 - n $.
%This way, one can equivalently define $ \mathbf{K}\left( x \right) $ (and all of the other above variants) when $ x $ is an encoded object instead of a string.

%% imported from paper 11

\subsection{Theoretical results}

The main idea to achieve the following theoretical results is to construct a randomly generated program that receives a formal theory (which contains all the computable procedures and statistical criteria in assembly theory) as input.
Then, it searches for a molecule (or object in an assembly space) with MA sufficiently high so as to make the pathway probability of spontaneous formation be sufficiently lower than the very own deceiving program's algorithmic probability, so that the divergence between these two probability distributions become statistically significant according to the arbitrarily chosen statistical method and significance level.

\begin{lemma}\label{SupInfthmLargeMA}
	Let $ \mathcal{ S } $ be infinite computably enumerable.
	Let $ \mathbf{F} $ be an arbitrary formal theory that contains assembly theory, including all the decidable procedures of
	%	the chosen statistical method, 
	the chosen method for calculating the assembly index (or approximating MA) of an object for a nested subspace of $ \mathcal{ S } $,
	and the program that decides whether or not the criteria for building the assembly spaces are met.
	Let $ k \in \mathbb{N} $ be an arbitrarily large natural number.
	Then, there are 
	a program $ \mathrm{ p }_y $, 
	$ \Gamma \subset \mathcal{ S } $ and $ y \in V\left( \Gamma \right) $ such that 
	\begin{equation}\label{SupInfeqLargeMA1}
		\mathbf{K}\left( y \right) + k 
		\leq
		\left| \mathrm{ p }_y \right| + k + \mathbf{O}(1)
		\leq
		c_\Gamma\left( y \right)
		\text{ ,}
		%		\mathbf{K}\left( y \right) \leq \mathbf{K}_\mathcal{ S }\left( y \right) + \mathbf{O}(1) + k \leq c_\Gamma\left( y \right)
		%		\text{ ,}
	\end{equation}
	where the function $ \; c_\Gamma \colon \Gamma \subset \mathcal{ S } \to \mathbb{N} \; $ gives the MA of the object $ y $ in the assembly space $ \Gamma $ (or $ \mathcal{ S } $)
	and
	$ \mathbf{U}\left( \mathrm{ p }_y \right) = y $.

	\begin{proof}
		Let $ p $ be a bit string that represents an algorithm running on a prefix universal Turing machine $ \mathbf{U} $ that receives $ \mathbf{F} $ and $ k $ as inputs.
		Then, it calculates $ \left| p \right| + \left| \mathbf{F} \right| + \mathbf{O}\left( \log_2\left( k \right) \right) + k $ and enumerates $ \mathcal{ S } $ while calculating $ c_\Gamma\left( x \right) $ of the object (or vertex) $ x \in V\left( \Gamma \right) \subset V\left( \mathcal{ S } \right) $ at each step of this enumeration.
		Finally, the algorithm returns the first object $ y \in  V\left( \mathcal{ S } \right) $ for which 
		\begin{equation}\label{SupInfeqProofLargeMA}
			\left| p \right| + \left| \mathbf{F} \right| + \mathbf{O}\left( \log_2\left( k \right) \right) + k + \mathbf{O}(1) \leq c_\Gamma\left( y \right)
		\end{equation}
		holds.
		In order to demonstrate that $ p $ always halts, just note that $ \mathcal{ S } $ is infinite computably enumerable.
        {Also, for any value of $ c_{ \Gamma' }\left( z \right) $ for some $ z \in V\left( \Gamma' \right) \subset V\left( \mathcal{ S } \right)  $, there is only a finite number of minimum rooted assembly subspaces (starting on any object in $ B_\mathcal{ S } $ and ending on $ z $) whose augmented cardinality is $ c_{ \Gamma' }\left( z \right) $,
		where $ B_\mathcal{ S } $ is the basis (i.e., the finite set of basic building blocks \cite{cronin}) of $ \mathcal{ S } $.}
		This implies that there is an infinite number of distinct values of $ c_{ \Gamma' }\left( z \right) $.
		Now, let $ \mathrm{ p }_y = \left< k , \mathbf{F} , p \right> $. 
		Finally, from Equation~\ref{SupInfeqProofLargeMA} and basic properties in AIT, we have it that
		\begin{equation}\label{SupInfeqProofLargeMA2}
			\begin{aligned}
				\mathbf{K}\left( y \right) + k
				\leq
				\left| \mathrm{ p }_y \right| + k + \mathbf{O}(1)
				\leq 
				\left| p \right| + \left| \mathbf{F} \right| + \mathbf{O}\left( \log_2\left( k \right) \right) + k + \mathbf{O}(1) \leq c_\Gamma\left( y \right)
				%				\text{ .}
			\end{aligned}
		\end{equation}
		holds for some sufficiently large $ k $.
	\end{proof}
\end{lemma}

\begin{lemma}\label{SupInfthmLowpathprobability}
	Let the conditions for Lemma~\ref{SupInfthmLargeMA} be satisfied.
	Let 
	\begin{equation*}
		\begin{aligned}
			\mathbf{ P } 
			\colon 
			\left\{ 
			%		\gamma = z \dots y \, \middle\vert z \in B_\mathcal{ S } , \, y \in V\left( \mathcal{ S } \right) 
			\Gamma' \middle\vert \exists x \left( c_\Gamma\left( y \right) = x \right) , \, \Gamma' \subseteq \Gamma \subset \mathcal{ S } \text{ is rooted}, \text{ and } y \in V\left( \Gamma' \right)
			\right\} 
			\to 
			\left[ 0 , 1 \right]
		\end{aligned}
	\end{equation*}
	be an arbitrary probability measure on the set of pathways in $ \mathcal{ S } $ and $ \mathrm{ p }_\mathbf{ P } $ a program that computes a computable function that outputs an upper bound for $ \mathbf{ P } $ such that for every $ \epsilon' > 0 $ and $ \Gamma \subset \mathcal{ S } $, there are $ \Gamma' \subset \mathcal{ S } $ and $ x \in V\left( \Gamma' \right) $ with $ \Gamma \subseteq \Gamma' $ and  $ \mathbf{ P }\left( { \Gamma^* }_{ x } \right) \leq \mathbf{U}\left( \left< { \Gamma^* }_{ x } , \mathrm{ p }_\mathbf{ P } \right> \right) < \epsilon' $.
	%	Let $ k' \in \mathbb{N} $ be an arbitrary natural number.
	Let $ 1 \geq \epsilon > 0 $ be an arbitrary encodable real number.
	Let $ k \in \mathbb{N} $ be an arbitrarily large natural number.
	Then, there are a program $ \mathrm{ p }_\epsilon $,
	$ \Gamma \subset \mathcal{ S } $, and $ y \in V\left( \Gamma \right) $
	such that Lemma~\ref{SupInfthmLargeMA} is satisfied with $ y $ and
	\begin{equation}\label{SupInfeqLowprobability1}
		\mathbf{K}\left( y \right) + k 
		\leq
		\left| \mathrm{ p }_\epsilon \right| + k + \mathbf{O}(1) 
		\leq
		c_\Gamma\left( y \right)
		\text{ ,}
		%		\mathbf{K}\left( y \right) \leq \mathbf{K}_\mathcal{ S }\left( y \right) + \mathbf{O}(1) + k + \mathbf{O}(1) \leq c_\Gamma\left( y \right)
		%		\text{ ,}
	\end{equation}
	and
	\begin{equation}
		\begin{aligned}
			\mathbf{ P }\left( { \Gamma^* }_{ y } \right) \leq \mathbf{U}\left( \left< { \Gamma^* }_{ y } , \mathrm{ p }_\mathbf{ P } \right> \right) < \epsilon
		\end{aligned}
	\end{equation}
	hold, where $ \mathbf{U}\left( \mathrm{ p }_\epsilon  \right) = { \Gamma^* }_{ y } $.
	
	\begin{proof}
		Let $ p' $ be a bit string that represents an algorithm running on a prefix universal Turing machine $ \mathbf{U} $ that receives $ \mathrm{ p }_\mathbf{ P } $, $ \epsilon $ and $ k $ as inputs.
		Then, 
		%it builds $ \mathrm{ p }_z = \left< k , \mathbf{F} , p \right> $ as in the proof of Lemma~\ref{SupInfthmLargeMA}.
		it enumerates the assembly pathways $ { \Gamma^* }_{ x } $ in $ \mathcal{ S } $ such that $ \left| p \right| + \left| \mathbf{F} \right| + \mathbf{O}\left( \log_2\left( k \right) \right) + \left| p' \right| + \left| \mathrm{ p }_\mathbf{ P } \right| + \mathbf{O}\left( \log_2\left( \epsilon \right) \right) + k + \mathbf{O}(1) \leq c_{ \Gamma }\left( x \right) $,
		%		and
		%		$ c_{ \Gamma' }\left( z \right) \leq c_{ \Gamma' }\left( x \right) $
		%		for 
		%		$ \Gamma' \subset \mathcal{ S } $ such that $ z \in V\left( \Gamma \right) \subseteq V\left( \Gamma' \right) $, 
		$  \mathbf{U}\left( \left< { \Gamma^* }_{ x } , \mathrm{ p }_\mathbf{ P } \right> \right) < \epsilon $,
		and Lemma~\ref{SupInfthmLargeMA} holds for $ x $ given $ k $. 
		Finally, it returns this first $ { \Gamma^* }_{ x } $ in this enumeration.
		Now, let $ \mathrm{ p }_\epsilon = \left< \mathrm{ p }_\mathbf{ P } ,  \epsilon, k , p' \right> $.
		Therefore,
		in addition to the arguments in the proof of Lemma~\ref{SupInfthmLargeMA}, the desired theorem follows from the fact that program $ p' $ always halts because of our initial assumptions on program $ \mathrm{ p }_\mathbf{ P } $ and the probability distribution given by $ \mathbf{ P }  $.
	\end{proof}
\end{lemma}

\begin{lemma}\label{SupInfthmDeceivingarbitraryprogram}
	Let the conditions for Lemmas~\ref{SupInfthmLargeMA} and~\ref{SupInfthmLowpathprobability} be satisfied.
	Let $ \mathbf{F}' \supseteq \mathbf{F} $ be a formal theory that also includes the chosen statistical method, the criteria for the arbitrarily chosen statistical significance level, the chosen computable method for approximating $ \mathbf{ P }  $ from above with program $ \mathrm{ p }_\mathbf{ P } $,
	and the acceptable maximum error $ \mathscr{ E } \in \mathbb{N} $ for an overestimation of the complexity of an object in $ \mathcal{ S } $.
	Then, there are a program $ \mathrm{ p }_d $,
	$ \Gamma \subset \mathcal{ S } $, and $ y \in V\left( \Gamma \right) $
	such that 
	Lemma~\ref{SupInfthmLowpathprobability} is satisfied with $ y $
	and $ \mathbf{F}' $ decides that the divergence of the (sub-)universal distribution
	%	$
	%%	\begin{equation}
		%%		\begin{aligned}
			%%			\sum\limits_{ \left| p \right| \geq \left| \mathrm{ p }_\epsilon \right| , \, p \in \mathbf{ L } }
			%%			\frac{ 1 }{ 
				%				2^{ - \left| \mathrm{ p }_d \right| } 
				%%			}
			%%		\end{aligned}
		%%	\end{equation}
	%	$
	from 
	$ \mathbf{ P } $ is statistically significant,
	where $ \mathbf{U}\left( \mathrm{ p }_d  \right) = { \Gamma^* }_{ y } $ and $ \left| \mathrm{ p }_d \right| + \mathscr{ E }
	<
	c_\Gamma\left( y \right) $ hold.

	\begin{proof}
		Let $ \mathrm{ p }_d $ be a bit string that represents an algorithm running on a prefix universal Turing machine $ \mathbf{U} $ that includes the computation of $ \mathrm{ p }_k $ and $ \mathrm{ p }_\epsilon  $ (which are programs defined in the proofs of Lemmas~\ref{SupInfthmLargeMA} and~\ref{SupInfthmLowpathprobability}) as subroutines.
		Then, it searches for the first $ { \Gamma^* }_{ x } $, sufficiently small value of $ \epsilon $, and sufficiently large value of $ k \geq \mathscr{ E } + \mathbf{O}(1) $ such that Lemmas~\ref{SupInfthmLargeMA} and~\ref{SupInfthmLowpathprobability} are satisfied with $ x $,
		$ 
		\left| \mathrm{ p }_d \right| + \mathscr{ E }
		<
		\left| p \right| + \left| \mathbf{F}' \right| + \mathbf{O}\left( \log_2\left( k \right) \right) + \left| p' \right| + \left| \mathrm{ p }_\mathbf{ P } \right| + \mathbf{O}\left( \log_2\left( \epsilon \right) \right) + k + \mathbf{O}(1) 
		<
		c_\Gamma\left( x \right)
		%		\text{ ,}
		%		\end{equation}
	$ holds,
	and
	the divergence of $ 2^{ - \left| \mathrm{ p }_d \right| }  $ from 
	$ \mathbf{U}\left( \left< { \Gamma^* }_{ x } , \mathrm{ p }_\mathbf{ P } \right> \right) $ is statistically significant according to the formal theory $ \mathbf{F}' $.
	Finally, the algorithm returns this first assembly pathway $ { \Gamma^* }_{ x } $ as output.
	Note that, since the value of $ 2^{ - \left| \mathrm{ p }_d \right| }  $ is fixed, one can always employ program $ \mathrm{ p }_\mathbf{ P } $ and the statistical criteria in theory $ \mathbf{F}' $ to find an arbitrarily lower probability than $ 2^{ - \left| \mathrm{ p }_d \right| }  $ so that the resulting probability distribution (i.e., the probability distribution given by $ \mathbf{ P } $) diverges from the sub-universal (a priori probability) distribution. 
	This holds because: of the algorithmic coding theorem, which implies that $ 2^{ - \left| \mathrm{ p }_d \right| }  $ is a lower bound for the sub-universal a priori probability upon language $ \mathbf{L}' $, where $ \mathrm{ p }_d \in \mathbf{L}' \subseteq \mathbf{L} $;
	%		the algorithmic probability of $ { \Gamma^* }_{ y } $ (except for an independent constant);
	and of the fact that $ \mathrm{ p }_\mathbf{ P } $ is a program that computes a (computable) function that outputs an upper bound for $ \mathbf{ P } $.\footnote{ Also note that $ \mathbf{F}' $ does not actually need to be able to compute the value of the sub-universal a priori probability of $ x $ because one already knows $ 2^{ - \left| \mathrm{ p }_d \right| }  $ is a lower bound for it and $ \mathbf{U}\left( \left< { \Gamma^* }_{ x } , \mathrm{ p }_\mathbf{ P } \right> \right) $ is an upper bound for the \emph{optimal} pathway probability of $ x $.}
	Additionally, this divergence eventually becomes statistically significant (as the divergence increases) because the probability of occurrence of a sequence of events following an empirical probability distribution, which diverges from the original distribution that the events are generated, eventually decreases as the divergence sufficiently increases. 
	Also note that $ \left| \mathrm{ p }_d \right| \leq \left| \mathrm{ p }_k \right| + \left| \mathrm{ p }_\epsilon \right| + \mathbf{O}(1) $.
	Therefore, since $ k $ and $ \epsilon $ were arbitrary in Lemma~\ref{SupInfthmLowpathprobability} and all the statistical methods in $ \mathbf{F}' $ are decidable by assumption, we have that $ \mathrm{ p }_d $ always halts.
\end{proof}
\end{lemma}

\begin{theorem}\label{SupInfthmDeceivingmolecule}
Let the conditions for Lemma~\ref{SupInfthmDeceivingarbitraryprogram} be satisfied.
Let $ \mathcal{ S } $ be an infinite assembly space whose set of randomly generated (computable) generative processes include one that can effect at least the computation of program $ \mathrm{ p }_d $, where Lemma~\ref{SupInfthmDeceivingarbitraryprogram} holds for $ \mathrm{ p }_d $ and $ y \in V\left( \mathcal{ S } \right) $.
Then:
\begin{itemize}
	\item the complexity error is larger than $ \mathscr{ E } $ (except for an independent constant);
	
	\item  and this error implies a statistically significant (according to $ \mathbf{F}' $) distinct frequency of occurrence of $ y $ than it was expected from the chosen assembly theory.
\end{itemize}

\begin{proof}
	From Lemma~\ref{SupInfthmDeceivingarbitraryprogram}, we have it that $ \mathbf{U}\left( \mathrm{ p }_d  \right) = { \Gamma^* }_{ y } $.
	Thus, from our assumptions and the definition of $ \mathbf{K}_\mathcal{ S } $, we have it that $ \mathbf{K}_\mathcal{ S }\left( y \right) \leq \left| \mathrm{ p }_d \right| + \mathbf{O}(1) $,
	which proves that the complexity error is larger than $ \mathscr{ E } $ from Lemma~\ref{SupInfthmDeceivingarbitraryprogram}.
	We also have that the probability of an assembly pathway being constructed by a randomly generated computable process is given by the sub-universal a priori probability of $ { \Gamma^* }_{ y } $ upon language $ \mathbf{L}_\mathcal{ S } $, i.e.,
	\begin{equation}
		\sum\limits_{
			\begin{array}{c}
				\mathbf{U}\left( p \right) = { \Gamma^* }_{ y }; \\
				p \in \mathbf{L}_\mathcal{ S }
		\end{array}  } 
		C \, \frac{ 1 }{ 2^{ \left| p \right| } } 
		\text{ .}
	\end{equation}
	Therefore, by replacing $ \mathbf{L}' $ with $ \mathbf{L}_\mathcal{ S } $ in the proof of Lemma~\ref{SupInfthmDeceivingarbitraryprogram}, we achieve a statistically significant (according to $ \mathbf{F}' $) distinct frequency of occurrence of $ y $ than it was expected from the chosen assembly theory.
\end{proof}
\end{theorem}

\begin{corollary}\label{SupInfthmDeceivingmoleculeinauniversalspace}
Let the conditions for Lemma~\ref{SupInfthmDeceivingarbitraryprogram} be satisfied.
Let $ \mathcal{ S } $ be an infinite assembly space whose randomly generated (computable) generative processes are capable of universal computation.
Then:
\begin{itemize}
	\item the complexity error is larger than $ \mathscr{ E } $ (except for an independent constant);
	
	\item  and this error implies a statistically significant (according to $ \mathbf{F}' $) distinct frequency of occurrence of $ y $ than it was expected from the chosen assembly theory.
\end{itemize}

\begin{proof}
	The proof follows directly from the fact that 
	$ \mathbf{K}\left( y \right) \leq \mathbf{K}_f\left( y \right) + \mathbf{O}(1) $ and from replacing $ \mathbf{L}' $ with $ \mathbf{L} $ in the proof of Lemma~\ref{SupInfthmDeceivingarbitraryprogram}.
\end{proof}
\end{corollary}

%As shown by Corollary~\ref{SupInfthmDeceivingmoleculeinauniversalspace}, the deceiving phenomena can be equally bad or even worse in case the molecules are byproducts of complex systems that are somehow capable of universal computation.

%\section{Complexity distortions in isomorphic transformations between acyclic multigraphs}\label{sectionSupInfDistortions} 

%\newpage
\section{Empirical results}\label{sectionSupInf-Results}

In 2017~\cite{zenilchem}, we explored the question of molecular and chemical separation using several data inputs and methods were we founded we could classify correctly organic from inorganic compounds using a standard chemical database of more than 15,000 compounds (as opposed to only 100 as in the Assembly Theory paper~\cite{cronin}.) We were able to successfully separate organic from inorganic compounds using not only chemical nomenclature names (InChI) but also structural bond network distance matrices found in the ChemicalData %[] \fsa{ Please, check this square brackets.}
(PubChem) repository available in Wolfram Mathematica.

In the early 2010s, we introduced a measure called Block Decomposition Method of which we proved most of its features in~\cite{bdm2} demonstrating that in the worse case without updating the underlying (universal) distribution based on the principles of algorithmic complexity, it would converge to Shannon Entropy but in the average case it would combine the power of what we called the Coding Theorem Method~\cite{ctm} with traditional information theory calculating local estimations of algorithmic complexity to find local patches of causality. The Block Decomposition Method looks for identical repeated blocks counting their abundance but also takes into account blocks that may have been assembled/produced by the same generating underlying mechanism, hence taking the concept of abundance of patterns beyond trivial statistical repetitions. In multiple applications, we connected it to molecular and chemical complexity~\cite{zenilchem}, behavioural complexity~\cite{chaos,plos}, cell biology~\cite{bdm}, molecular biology~\cite{nar}, causality~\cite{nmi,Zenil2020cnat,aidbook} and selection and evolution~\cite{santiago}.

Despite their statistical limitations, RLE, LZ and Huffman's coding algorithms are among the simplest coding algorithms introduced in the 1960s and 1970s. 
They are known not to be optimal for statistical or algorithmic compression, but optimal at doing what they were intended to do, that is, counting statistical copies in the form of minimum code lengths and approaching Shannon Entropy rate in the limit.

\subsection{Molecular Assembly classification exhibits similar or lower performance than existing statistical algorithms on multiple data types}

We compared the performance of  `Molecular Assembly' (MA) with measures of statistical (statistical compression) and algorithmic (BDM) nature under the four mass spectroscopy (MS) categories seen in~\cite[Figures 2-4]{cronin}. Note, that mass spectroscopy can also be referred to as mass spectrometry, and is used interchangeably. 

Various measures such as BDM, LZW, and RLE were first computed on the InChI code strings to show that other data inputs were able to classify the same molecules/compounds in~\cite{marshall_murray_cronin_2017}, followed by an application to bond connectivity matrices of structural chemical networks. Finally, we applied all these indexes to the 2D mass spectra matrices of the molecules or chemical systems in~\cite[Figures 2-4]{cronin}. 
To this end, we binarised the 2D mass spectrometry matrices provided in their supplementary data according to our methodology as described in Section~\ref{sectionSupInf-Methods}, and computed the rest of the values for all other measures. We then used correlation analysis and a combination of simple statistical tests to compare MA to the other indexes in classifying the chemical systems and molecules from Marshall's results in \cite[Figures 1-4]{cronin}, as life \textit{vs}$.$ non-life categories. The T- test and Kolmogorov-Smirnov (KS) test have been used for this purpose.  Unlike the t-test statistic, the Kolmogorov-Smirnov test provides a non-parametric goodness-of-fit test, assuming the data does not come from a Gaussian (Normal) distribution. It should be noted that while KS-tests by themselves are not classification algorithms, they are useful as part of a classification pipeline by providing a statistical measure of the similarity between sample distributions. A combination of the KS-test p-value along with other metrics like corrected t-test p-values and R-squared values from the correlation analysis allows for a more determination of whether two distributions are statistically distinguishable, and to evaluate the categories used in the figures in~\cite{cronin}. While other types of statistical tests are possible, in general it is not the case that one test will contradict other. Even conceding that these other measures do not necessarily outperform Molecular Assembly (MA) or the assembly index (which according to our statistics they do), this work still suggests all these other measures produce the same results and should have been explored in a comparison analysis in the original study~\cite{marshall_murray_cronin_2017} for good scientific practice. Had it been done, the authors may have learned more about their own index and may have placed it in the right context, cite previous literature, and provide a proportional coverage of its limited impact given its incremental, if anything, contribution.

Figure~\ref{SupInffigureTreediagrams} shows an illustration of the standard operation of Huffman coding as the first dictionary-based coding scheme on a typical example next to a popular example used by the authors of Assembly Theory~\cite{cronin}. In Figures~\ref{SupInffigureCompressioncorrelationplot}~and~\ref{figureLivingversusNonliving}, we test a set of statistical measures in preparation for the incorporation of mass spectroscopy data as done in~\cite{marshall_murray_cronin_2017}. Both 1D-RLE and 1D-Huffman coding schemes show a strong statistical correlation and linear correspondence with MA (see Fig.~\ref{SupInffigureCompressioncorrelationplot}). The one-dimensional and simplest lossless compression algorithms RLE and Huffman code compression lengths showed the strongest Pearson correlations with MA at R-values of 0.9001 and 0.896. The complete correlation analysis of the 114 molecules classification is provided in Table~\ref{SupInftableComplexityclassification}.

The results do not come as a surprise because these algorithms count repetitions even if they may do so in different ways or may represent data in a different fashion they still are able to pick the same repetitive signal independent of representation guaranteed by the principles of information theory as one cannot create or erase patterns by simply changing the tokens or the underlying vocabulary from a direct translation. All these results conform with the theoretical expectation of MA to be a compression scheme of the LZ family as proven in~\cite{abrahao2024}.

It has long been established that prediction is equivalent to lossless coding/compression and vice versa~\cite{Downey2010}. The compressed version of a phenomenon is a model that has to abstract its most salient properties. These results imply that these are not abstract disconnected properties of data but the properties that are most often the best and right explanation of a process. This is the main assumption of science. Therefore, compression is not only for data compression but is at the core of the practice of science, simulation, modelling, abstraction and prediction. Yet, the misunderstandings surrounding compression, Shannon Entropy, Kolmogorov complexity, Turing machines, computation universality, computability, evolutionary and development biology is deep and manifold~\cite{marshall_murray_cronin_2017,Marshall2019,cronin}

\subsection{Molecular classification by structural information in nomenclature codes}

Here we show how different data types classify the data in the same way as Molecular Assembly does, without access to other type of data or other algorithms different from statistical or compression/coding schemes as reported in~\cite{zenilchem} before moving to molecular bond distance matrices also reported in ~\cite{zenilchem} and mass spectral data as used in~\cite{marshall_murray_cronin_2017}.

InChI is an open standard identifier for chemical databases that facilitates effective identification of chemical compounds. The InChI algorithm converts input structural information into a unique identifier in a three-step process. A normalisation (to remove redundant information), canonicalisation (to generate a unique number label for each atom), and serialisation (to assign a string of characters) and as such, InChI codes include all the necessary information to uniquely map and build the structure of every chemical compound.

\begin{figure}
\centerline{\includegraphics[scale=0.11]{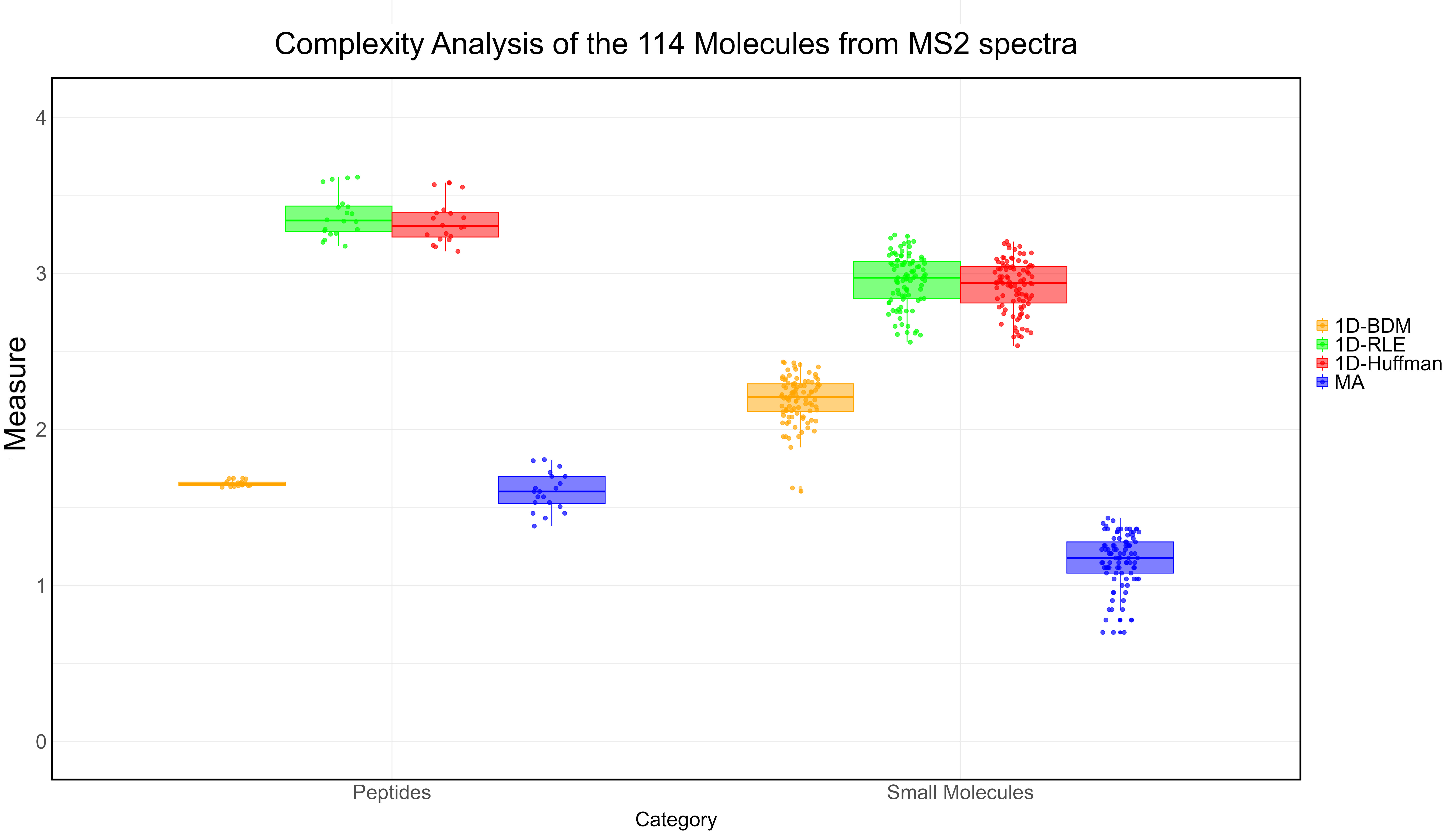}}\caption{\label{SupInffigureCompressioncorrelationplot}Correlation plot between the `Molecular Assembly' (MA) index (taken at face value, not recalculated only reclassified according to a larger category space as in~\cite{zenilchem}) and other compression scores on InChI codes as performed in~\cite{zenilchem} with the molecules in~\cite{marshall_murray_cronin_2017}. The vertical axes are the five complexity scores in log normalised scale for comparison purposes.}
 \end{figure}

The application of traditional statistical indexes analysis reveals the same separation as the reported in MA that also failed to cite the previous results in the field~\cite{zenilchem}. These popular statistical lossless compression algorithms are based on the same counting-copies principles used by Assembly Theory~\cite{cronin} and the Molecular Assembly~\cite{marshall_murray_cronin_2017}. 

As seen in Fig.~\ref{SupInffigureCompressioncorrelationplot}, the values of both 1D-RLE and 1D-Huffman coding show overlapping and nearly identical medians (horizontal line at centre) and ranges on the whisker plot. The plot shows that while the input data (mass spectral for MA versus InChI nomenclature for others), in all cases, a significant different signal across categories is found allowing each class to have different statistics for proper classification, in particular among natural and organic or inorganic.

 \begin{figure}[ht!]
\centerline{\includegraphics[scale=0.11]{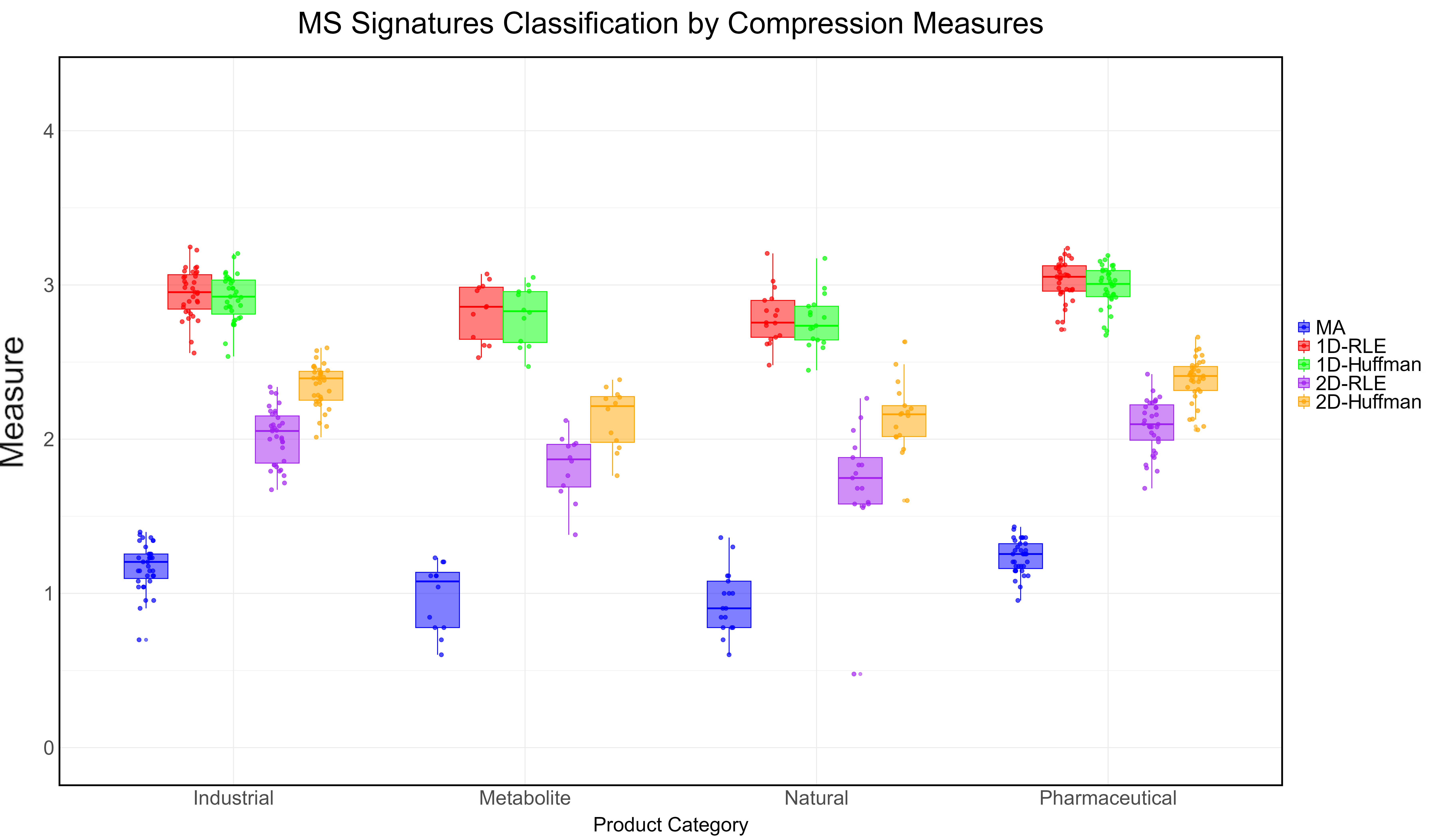}}
\caption{\label{SupInffigureCompressioncorrelationplot2}Same analysis of the application of multiple statistical indexes on the same set but according to the categories in~\cite[Figure 3]{cronin} showing the same separating properties. The strongest Pearson correlation was identified between 1D-BDM and the category of molecules (R= 0.828; P$<$0.0001).}
\end{figure}

\begin{table}[ht!]
	\begin{center}
		\begin{tabular}{c|c|c|c|c} 
			\textbf{Statistic} & \textbf{1D-BDM} & \textbf{1D-RLE}& \textbf{1D-Huffman}&\textbf{MA}\\
			\hline
			Pearson  &&&&\\
			Correlation (R) &  0.828 &  0.704 &  0.713 & 0.7111\\
			\hline
			99\% confidence &&&&\\
			interval &  0.87-0.76 &   0.59-0.78  &  0.61-0.793  &  0.60-0.79\\
			\hline
			R squared &  0.686 &  0.495 &  0.509 & 0.506\\
		\end{tabular}      \caption{\label{SupInftableComplexityclassification1}Table of Pearson Correlation values of MA and complexity indices across the two categories (small molecules and peptides) of the molecules in~\cite[Figure 3]{cronin}. BDM, RLE, and Huffman are given in log-normalised bits. As shown, BDM generates better statistics than MA without any adaptions or modifications, while Huffman shows near identical correlation performance as MA, thereby, supporting the findings.}
	\end{center}
\end{table}

The comparison of measures across the four categories of MS molecules is shown in Table~\ref{SupInftableComplexityclassification2} with respect to increasing the molecular weight (MW) to better visualise the trends across living and non-living bio-signatures. The Pearson correlation test was assessed on the various complexity and compression measures in relation to molecular weight (MW) with an alpha value of 0.01 (99 percent confidence interval) for which the one-tailed p-values were significant (P$<$0.0001) for all indexes compared in Table~\ref{SupInftableComplexityclassification2}. The one-tail P-value tests were performed instead of the two-tail tests since our (previous) analyses inferred a unidirectional linear relationship in the trend patterns. As shown in Table~\ref{SupInftableComplexityclassification2}, 1D-BDM had the highest Pearson correlation with MW (R=0.905), followed by LZW compression (R=0.9028). MA has a correlation score of 0.8055.

\subsection{Molecular classification by chemical bond distance matrices and spectral data}

We now test these measures on objects closer to the instrument measurement of the chemical data, including the mass spectral data used by Marshall et al. in~\cite{marshall_murray_cronin_2017}. The two-dimensional distance matrices of the mass spectroscopy (MS) data were binarised and converted using a threshold of three before being subjected to the compression algorithms. The 2D-RLE and 2D-Huffman code compression lengths obtained Spearman correlation values of 0.7967 and 0.7537, respectively with MA (the Pearson scores were comparable). The gzip compression showed a Spearman correlation of 0.804.

A strong Pearson correlation with an R-value of 0.8823 was observed between 1D-BDM and MA for the 99 molecules available in the MS data set (see Fig.~\ref{figureLivingversusNonliving}).  LZW compression shared a close Pearson's correlation score of 0.8738 with MA. All correlation measures obtained a statistically significant one-tailed p-value ($P<0.0001$). 

%\begin{figure}[ht!]
%	\centerline{\includegraphics[scale=0.11]{Figure_2_Chemical_space.png}}
%	\caption{\label{figureLivingversusNonliving}Classification of molecular complexity by multiple complexity indexes originally used to create the chemical space for the mass spectroscopy (MS) profiles (log-scale). All measures other than MA applied to bond molecular distance matrices some of which outperform MA and their mass spectra at distinguishing organic from non-organic molecules found in the MS dataset of the MA paper~\cite{marshall_murray_cronin_2017}, as demonstrated by greater separation and smaller variance results across the different complexity measures among the molecular subgroups. MA does not display any particular advantage when compared against proper control experiments, and performs similarly to the simplest of the statistical algorithms applied to all the tested data representations including molecular distance matrices (as shown here for all measures but MA) or the mass spectral data provided by the authors of Assembly Theory (shown on the plot from the authors' results that could not be fully reproduced due to lack of data made available in~\cite{cronin} but we took at face value) for comparison purposes.}
%\end{figure} 

%These findings indicate that the methods behind Assembly Theory, on which Molecular Assembly (MA) is based, can easily be replaced by one of the first and simplest compression schemes even in the case , 1D-RLE, in the classification of mass spectrometry (MS) complexity. 

The Molecular Assembly indices did not show any significant advantage when compared with other measures that were introduced several decades ago when computer compression algorithms where designed based upon the same modular statistical principle of repetition and modular counting re-introduced by MA. Nor were the MA indices able to show any particular advantage over indices that are non-computable but capable of being approximated from above and based on resource-bounded variants of algorithmic complexity (such as BDM~\cite{bdm,2d}), which the authors of MA disqualify a priori~\cite{cronin} without any evidence or control experiments, on account of their semi-computable nature (where `semi' means they can be approximated using various methods, as is the case, for example, with protein folding).

% \section{Comparison of correlation with molecular weight}

\begin{table}[h!]
	\begin{center}
		\begin{tabular}{c|l|l|l} % <-- Alignments: 1st column left, 2nd middle and 3rd right, with vertical lines in between
			\textbf{Statistic} & \textbf{MA} & \textbf{LZW}& \textbf{1D-BDM}\\
			\hline
			Pearson  Correlation (R) &  0.897 &  0.902 &  0.905\\
			\hline
			99\% confidence 
			interval &  0.832 - 0.938 &   0.84 - 0.941  &  0.845 - 0.943\\
			\hline
			R squared &  0.805 &  0.815 &  0.820\\
		\end{tabular}
		\caption{\label{SupInftableComplexityclassification2}Table of Pearson correlation values corresponding to Fig. 1. LZW and BDM are given in bits, meaning the length of a compressed description of the object, including the number of steps. Both LZW and BDM generate better separating statistics on InChI codes than MA without any adaptions or modifications.}
	\end{center}
\end{table}

The correlation analysis suggests a stronger positive linear relationship between MW and measures from algorithmic information dynamics, such as BDM and LZW, in contrast to that between MW and MA. As such, other indices on other chemically derived data representation can be better predictors of increasing molecular complexity in the MS signatures classification as reported before in~\cite{zenilchem}.

MA and Shannon Entropy had a similar statistical significance in classifying the mass spectroscopy (MS) data into their four distinct categories, with t-values of 15.96 and 20.96, respectively at df $=$ 100. The Kolmogorov-Smirnov distances were 0.828 and 1, respectively. 

The suggestion of Assembly Theory was that MA can predict living \textit{vs}$.$ nonliving molecules tested on a cherry-picked small subset of biological extracts, between abiotic factors, and inorganic (dead), as shown in~\cite[Figure 4]{cronin}. We repeated the experiment using the binarised MS2 spectra peaks matrices provided in the source data in~\cite{marshall_murray_cronin_2017}. 18 of the extracts and molecular MS2 spectra were obtained, ignoring the blinded samples shown in \cite[Figure 4]{cronin}. 
Our reproduced findings on their \cite[Figure 4]{cronin}, are shown in our Fig.~\ref{figureBiologicalComplexityplot}. 
By including the 114 molecules from~\cite[Figure 3]{cronin} with the 18 molecules of~\cite[Figure 4]{cronin}, we performed correlation analysis on all 132 signatures, with 5 categories: small molecules, peptides, abiotic, dead (inorganic, such as coal and quartz) and biological extracts (which includes yeast, E.coli, etc.). The Pearson correlation was strongest between 1D-BDM and the category (R= 0.951), followed by 1D-RLE and 1D-Huffman having a near-identical Pearson correlation of R= 0.843 and R= 0.842, respectively. MA has the poorest correlation with the categories, with a correlation of R= 0.448. All Pearson scores were statistically significant (P$<$0.0001). The results for all 132 molecules are shown in Table~\ref{SupInftableComplexityclassification}. Therefore, our findings collectively conclude that when considering both the mass spectrometry signatures of~\cite[Figures 2B, 3, and 4]{cronin}, together, the coding indexes systematically outperform MA index, as a discriminant of living \textit{vs}$.$ non-living systems.

\begin{table}[h!]
	\begin{center}
		\begin{tabular}{c|c|c|c|c} 
			\textbf{Statistics} & \textbf{1D-BDM} & \textbf{1D-RLE}& \textbf{1D-Huffman}&\textbf{MA}\\
			\hline
			Pearson  &&&&\\
			Correlation (R) &  0.95 &  0.84 &  0.84 & 0.45\\
			\hline
			99\% confidence &&&&\\
			interval &  0.96-0.93 &   0.88-0.78  &  0.88-0.78  &  0.57-0.3\\
			\hline
			R squared &  0.904 &  0.711 &  0.709 & 0.201\\
		\end{tabular}
		\caption{\label{SupInftableComplexityclassification}Table of Pearson Correlation values of MA and complexity measures across all 132 molecules, including the biological extracts from~\cite[Figure 4]{cronin}. BDM, RLE, and Huffman are given in log-normalized bits. As shown, BDM and the compression algorithms generate better statistics than MA.}
	\end{center}
\end{table}

There is a significant level of variance in the MA scores of these biological mixtures and extracts, as indicated in~\cite[Figure 4]{cronin}. Further, given that the MA or complexity of the biological extracts shown in~\cite[Figure 4]{cronin} are mixtures derived from the small molecules and peptides in the MA chemical space constructed from~\cite[Figure 3]{cronin} data, and by virtue of other coding indexes outperforming their chemical MA space classification, we can conclude that MA theory is sub-optimal and a limited subset of compression measures provided within the algorithmic complexity framework. 

This section completes the control experiments missing in~\cite{marshall_murray_cronin_2017,cronin,croninnature} where it is claimed novelty at `counting copies' as a measure or its application to a special type of data that is not needed to produce the same results (from any tested data).

\begin{table}[h!]
    \centering
    \begin{tabular}{lcc}
        \toprule
        Measure & Kruskal-Wallis p-value & ANOVA p-value \\
        \midrule
        1D-BDM & 0.02 & 0.01 \\
        MA & 0.00 & 0.00 \\
        1D-Huffman & 0.03 & 0.01 \\
        1D-RLE & 0.03 & 0.01 \\
        \bottomrule
    \end{tabular}
    \caption{The Kruskal-Wallis and ANOVA statistical significance test on Mass Spectrometry Data. Validation by Kruskal-Wallis and ANOVA tests along with one-tailed t-tests of the MS2 data analysis shown in Figure 2 in our main article (i.e., complexity analysis on Dead, Inorganic, and Biological samples) or Figure~\ref{figureBiologicalComplexityplot} can provide more robust and credible significance testing (p-values) by examining the data from multiple statistical perspectives. Kruskal-Wallis is a non-parametric alternative to ANOVA, and using both parametric and non-parametric tests can help validate the findings and rule out potential issues with normality assumptions. Both test results were computed using the \texttt{scipy.stats} package in Python.}
    \label{SupInftab:significance-tests}
\end{table}

\end{document}